\documentclass[aps,prl,twocolumn,floatfix]{revtex4-2}

\usepackage{graphics} 
\usepackage{color} 
\usepackage{bm} 
\usepackage{bbm}
\usepackage[dvipsnames]{xcolor,colortbl}
\usepackage{amsfonts}
\usepackage[normalem]{ulem}  

\definecolor{DarkYellow}{RGB}{80, 80, 0}

\usepackage{amsmath,amssymb,graphicx}
\usepackage{amsbsy}
\usepackage{latexsym}
\usepackage{xcolor}
\usepackage{graphicx}
\usepackage{psfrag}
\usepackage[normalem]{ulem}
\usepackage{bm}
\usepackage{hyperref}
\usepackage[SF,sf]{subfigure}
\hypersetup{colorlinks=true, linkcolor=red, citecolor=blue, urlcolor=blue}
\usepackage{float}
\usepackage{multirow}
\usepackage{caption}
\begin{document} 

\title{Navigating complex phase diagrams in soft matter systems}

\author{Michael Wassermair$^{1,2,3}$}
\author{Gerhard Kahl$^{1}$}
\author{Roland Roth$^{4}$} 
\author{Andrew J.~Archer$^{2}$}

\affiliation{$^1$Institut f\"ur Theoretische Physik, TU Wien, Wiedner Hauptstra{\ss}e 8-10, A-1040 Vienna, Austria}
\affiliation{$^2$Department of Mathematical Sciences and Interdisciplinary Centre for Mathematical Modelling, Loughborough University, Loughborough LE11 3TU, United Kingdom}
\affiliation{$^3$Institute of Science and Technology Austria, 3400 Klosterneuburg, Austria}
\affiliation{$^4$Institute for Theoretical Physics, University of T\"ubingen, D-72076 T\"ubingen, Germany}


\date{\today}

\begin{abstract}
Colloidal fluids can exhibit complex phase behavior and determining phase diagrams via experiments or computer simulations can be laborious. We demonstrate that the dispersion relation $\omega(k)$, obtained from dynamical density functional theory for the uniform density system, is a highly versatile tool for {\it predicting} where in the phase diagram complex crystals form. The sign of $\omega(k)$ determines whether density modes with wavenumber $k$ grow or decay over time. We demonstrate the predictive power by investigating the complex phase behavior of particles interacting via core-shoulder pair potentials. With complementary Monte Carlo simulations, we show that regions of the phase diagram where $\omega(k)$ has one or several unstable (growing) wavenumbers are also where crystalline phases occur. Going further, by tuning these unstable wavenumbers via the interaction-potential and state-point parameters, we design systems with quasicrystals in the phase diagram. We identify a system with a certain shoulder-range exhibiting at least 10 different phases. Our general approach accelerates considerably the mapping of complex phase diagrams, crucial for the design of new materials.
\end{abstract}

\maketitle

\newpage

Self-assembly into complex structures underpins much of soft-matter, materials and condensed-matter science.
Collections of particles with relatively simple interactions, such as attractive-wells, repulsive ramps and/or square shoulder systems, are capable of generating a variety of complex structures \cite{nagel2017experimental, barrat2023soft}.
A fascinating and useful property of soft-matter and colloidal systems is that the potentials between the particles can be tuned in a seemingly unlimited manner by the chemistry and geometry of the constituent molecules, especially those on the surfaces of the particles \cite{israelachvili2011intermolecular, liu2022controlling}.
For example, by grafting polymers onto the surface of colloidal spheres, core-shell particles are created that assemble into a wide variety of structures, determined by the polymer lengths and grafting density \cite{ghosh2012core, vogel2012ordered, rey2017anisotropic, Menath2021, Ickler2022}.
This control enables the formation of structures with designed symmetries, optical and other properties, for numerous applications \cite{li2021colloidal}.

Determining the phase diagram, specifically mapping where the solid crystalline, or even quasicrystalline (QC) phases arise, is crucial for designing new materials \cite{dijkstra2021predictive}.
This effort can be expensive and time-consuming to realize in experiments.
Computer simulations can be cheaper, but often still prolonged.
Thus, the predictive theory we develop, albeit one that does not give the full answer, to identify where the solid phases arise is very powerful. 
We show how a purely analytic, computationally cheap, theory for the growth or decay of density perturbations is able to guide Monte Carlo (MC) computer simulations to the relevant portions of the phase diagram, saving much effort.
Our theory is based on classical density functional theory (DFT) \cite{Evans1979, Evans_2016, hansen13} and its dynamical extension, DDFT \cite{hansen13, marconi1999dynamic, Archer2004, te2020classical}.
We demonstrate the effectiveness of our approach by applying it to systems of core-shell particles which (we show) can exhibit complex phase diagrams.
We present a two-dimensional (2D) example with at least 10 different phases.
We expect our general approach to become a widely applicable initial-survey tool for mapping soft-matter phase diagrams.

We demonstrate our approach for 2D systems interacting via the hard-core, square-shoulder (HCSS) pair potential $\phi(r)=\phi_\sigma(r)+\epsilon H(\lambda\sigma-r)$, where $\phi_\sigma(r)$ is the hard-disks potential with diameter $\sigma$ and $\epsilon H(\lambda\sigma-r)$, with the Heaviside function $H(r)$, is a square shoulder potential with range $\lambda \sigma$ and interaction strength $\epsilon > 0$.
The state of the system is determined by the dimensionless density $\rho_{\rm b}^\star = \rho_{\rm b} \sigma^2$ (with corresponding packing fraction $\eta = \pi \rho_{\rm b}^\star/4$), where $\rho_{\rm b}$ is the average bulk density, and dimensionless temperature $T^\star=k_{\rm B}T/\epsilon$, where $k_{\rm B}$ is Boltzmann's constant and $T$ is the temperature.
Henceforth, we set $\sigma=1$, $k_{\rm B}T = \beta^{-1} = 1$ and drop the $\star$ superscript, so that $T = 1/\epsilon$.

Despite the simplicity, HCSS-particles can self-assemble into a plethora of different phases involving single or multiple characteristic length scales and varying levels of complexity \cite{Glaser2007, Dobnikar_2008, Fornleitner_2008, Fornleitner_2010, Dotera2014, ziherl2016geometric, Dotera2017, pattabhiraman2017formation, Menath2021}.
There is no gas-liquid phase transition, but on increasing $\eta$ at sufficiently low $T$, one typically observes cluster, stripe and hole phases, with details depending greatly on $\lambda$.
Many other different phases can also be observed either instead or as well, including quasicrystals (QCs) \cite{Dotera2014, ziherl2016geometric, Dotera2017}.

The importance of identifying the wavelength $\ell$ of the periodic modulations in the equilibrium density distribution $\rho_0(\mathbf{r})$ of crystalline structures has long been understood \cite{alexander1978should}.
Building on such ideas, soft pair potentials have been designed so that the particles self-assemble into a variety of QCs and other structures \cite{barkan2011stability, archer2013quasicrystalline, barkan2014controlled, archer2015soft,  savitz2018multiple, Ratliff2019, subramanian2021density, Archer2022}.
QCs arise from the coupling of density modes with wavenumbers $k_1=2\pi/\ell_1$ and $k_2=2\pi/\ell_2$, with ratio $k_1/k_2$ taking specific values \cite{lifshitz1997theoretical}.
It was also observed that the liquid state static structure factor $S(k)$ \cite{hansen13} can be used to identify the specific $k_i$ that, when the system is cooled below the freezing point, govern the crystalline phases \cite{wassermair2024fingerprints}.
These ideas are the background to the present theory that enables navigation of complex phase diagrams and provides a basis for designing interaction potentials that give rise to specified self-assembled structures.

The start-point is to analyze the non-equilibrium dynamics of a homogeneous bulk liquid with density $\rho_{\rm b}$, following a quench to a certain temperature $T$.
If this quench is to a state point where the liquid is unstable, the density distribution becomes heterogeneous $\rho({\bf r},t)$, evolving in time $t$ towards the equilibrium $\rho_0({\bf r})$, governed by the continuity equation
$\frac{{\partial \rho}}{\partial t}= - \nabla \cdot {\bf j}$,
where ${\bf j}(\mathbf{r},t)$ is the current.
DDFT \cite{marconi1999dynamic, Archer2004, hansen13, te2020classical} provides a good approximation for ${\bf j}$, as long as a reliable approximation is available for the grand potential functional
\begin{equation}\label{eq:grand_pot}
\Omega[\rho]= F_{\rm id}[\rho]+F_{\rm ex}[\rho]+\int\rho(\mathbf{r})(V_{\rm ext}(\mathbf{r})-\mu){\rm d}\mathbf{r},
\end{equation}
where $F_{\rm id}[\rho]$ is the ideal-gas contribution, $F_{\rm ex}[\rho]$ is the excess Helmholtz free energy that depends on the nature of the particle interactions, $V_{\rm ext}(\mathbf{r})$ is the external potential (here assumed to be zero) and $\mu$ is the chemical potential \cite{hansen13, Evans1979}.
For particles with overdamped stochastic (Brownian) equations of motion, DDFT gives
\begin{equation}\label{eq:j_DDFT}
{\bf j}=\Gamma\rho\nabla\frac{\delta\Omega[\rho]}{\delta\rho},
\end{equation}
where $\Gamma=D/k_{\rm B}T$ is the mobility coefficient, with $D$ the diffusion coefficient  \cite{marconi1999dynamic, Archer2004, hansen13, te2020classical}.
For dense atomic or molecular fluids, where the particles evolve under Newton's equations of motion, it turns out that \eqref{eq:j_DDFT} still provide a reasonable description \cite{archer2006dynamical, archer2009dynamical}, as long as the system is not too far from equilibrium.
The density distribution of the quenched liquid is never perfectly uniform; it must have small amplitude random density perturbations ${\tilde \rho}({\bf r},t)\equiv\rho({\bf r},t)-\rho_{\rm b}$ around the average value, $\rho_{\rm b}$.
To investigate the initial growth/decay dynamics of such perturbations after the quench, it suffices to consider the linearized form of the continuity equation, $\frac{{\partial \tilde \rho}}{\partial t}={\mathcal L}{\tilde \rho}+{\mathcal O}({\tilde \rho}^2)$.
Here, ${\mathcal L}$ is a linear operator obtained from performing a functional Taylor expansion of Eq.~\eqref{eq:j_DDFT} in powers of $\tilde{\rho}$.
We obtain ${\mathcal L} = D \nabla^2 -D \rho_{\rm b}\nabla^2c^{(2)} \otimes$, with $\otimes$ denoting convolution, i.e.\ $c^{(2)}\otimes{\tilde \rho}\equiv\int d{\bf r}'c^{(2)}({\bf r}-{\bf r}'){\tilde \rho}({\bf r}',t)$, and where $c^{(2)}\equiv-\beta\delta^2F_{\rm ex}/\delta\rho^2$ is the pair direct correlation function (pDCF) \cite{Archer2004, hansen13}.
In the liquid, the pDCF may instead be obtained via the Ornstein-Zernike equation \cite{hansen13}.
We introduce the dispersion relation $\omega(k)$ by decomposing the perturbation ${\tilde \rho}$ into its Fourier components
\begin{equation}\label{eq:rho_modes}
{\tilde \rho}({\bf r},t)=\sum_{\bf k}{\hat \rho}_{\bf k}e^{i{\bf k}\cdot{\bf r}+\omega(k)t},
\end{equation}
where $k\equiv|{\bf k}|$, ${\hat \rho}_{\bf k}$ are the ($t=0$) initial amplitudes and thus (depending on the sign) $\omega(k)$ is the growth/decay rate of each mode (note our definition of $\omega(k)$ following e.g.\ \cite{dee1983propagating, archer2012solidification}, {\em without} prefactor $i=\sqrt{-1}$). 
From the linearized dynamics, the dispersion relation is obtained in terms of the Fourier transform of the pDCF ${\hat c}(k)$ as
\begin{equation}\label{eq:omega}
\omega(k)=-Dk^2[1-\rho_{\rm b}{\hat c}(k)].
\end{equation}
In an equilibrium liquid $S(k)=1/[1-\rho_{\rm b}{\hat c}(k)]$, so in this case $\omega(k)=-Dk^2/S(k)\leq0$, for all $k$.
In regions of the phase diagram where solid phases arise (where the liquid is either unstable or metastable) we obtain ${\hat c}(k)$ {\em generated} by DFT.
This relies on having a suitable approximation for $F_{\rm ex}$.
However, as we show, even with an approximate functional that does not always distinguish correctly {\em all} the different crystal phases, we still find that the resulting $\omega(k)$ reliably estimates the locations in the phase diagram where the solid phases arise.
The approximation for $F_{\rm ex}$ used here is a standard mean-field approximation, splitting it into a hard-disk part, treated using fundamental measure theory (FMT) \cite{Roth2012, Roth2010}, plus a random-phase-approximation (RPA) for the shoulder part.
Further details are in the supplementary information (SI) below, including an analytic expression for $\hat{c}(k)$ \cite{Thorneywork2018}.
Note that $\omega(k)$ effectively depends solely on thermodynamic structural quantities, therefore our analysis is independent of the assumed dynamics (Brownian or Newtonian).

\begin{figure}[t]
  \centering
\includegraphics[width=1.0\columnwidth]{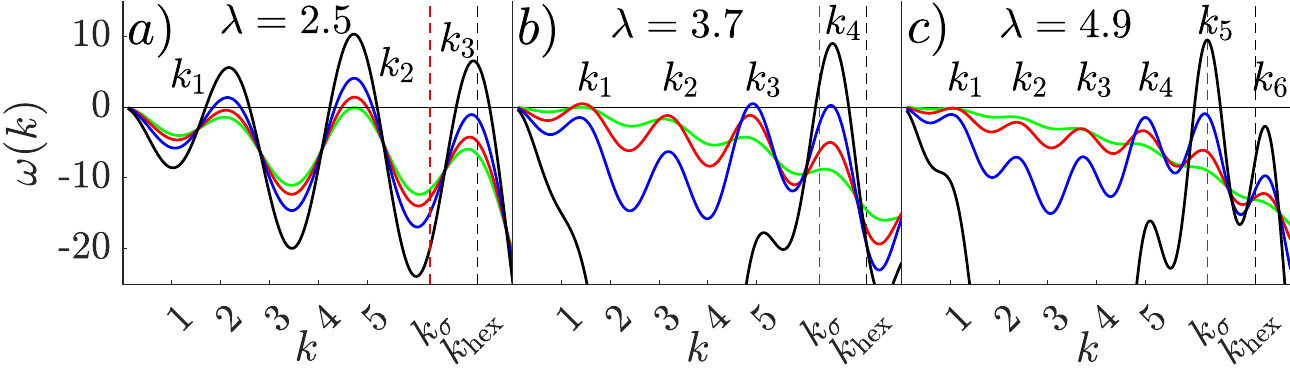}
\includegraphics[width=1.0\columnwidth]{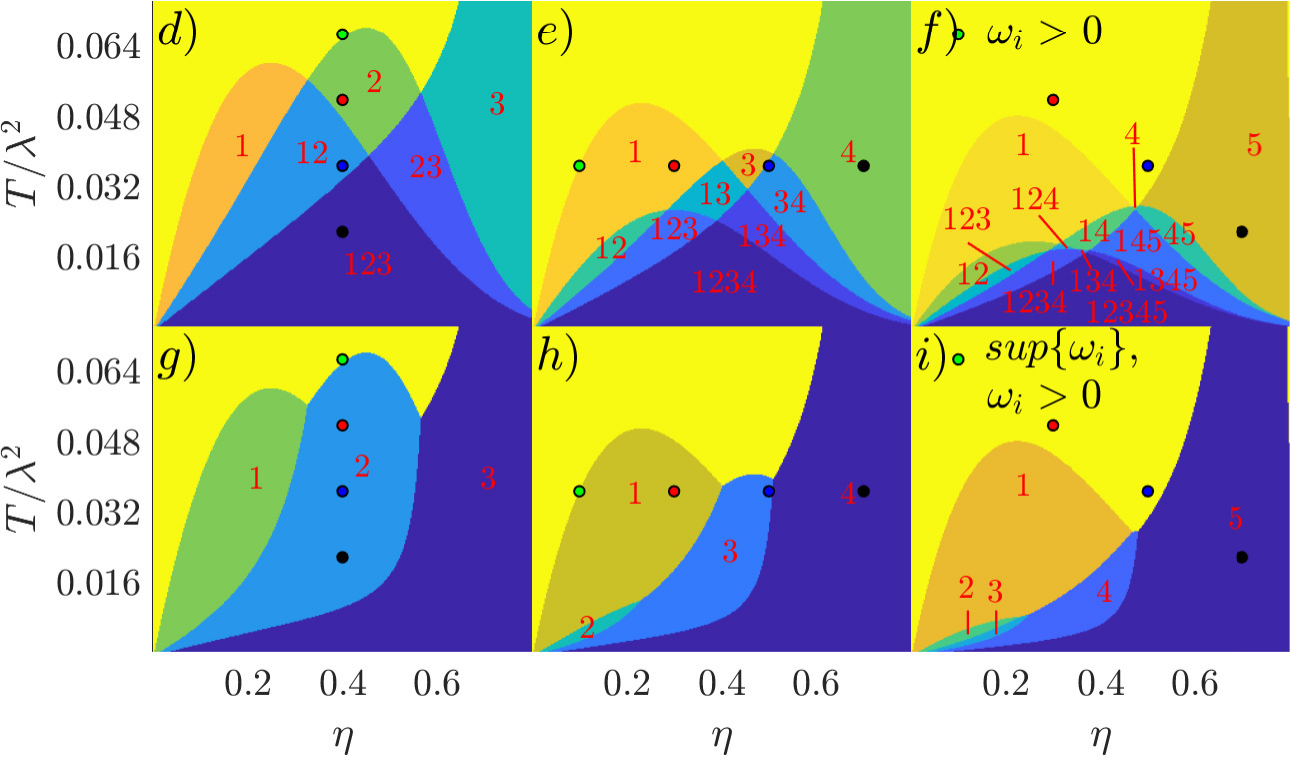}
\caption{Top row: dispersion relation $\omega(k)$ for the HCSS-system for a) $\lambda=2.5$, b) $\lambda=3.7$ and c) $\lambda=4.9$.
The maxima at $k_i$ (with $i=1,2,3,\dots$) are indicated. 
The state points at which $\omega(k)$ is evaluated are the correspondingly colored dots in the $(T/\lambda^2,\eta)$-plane diagrams below each, labeled d) to i).
The red and black vertical dashed lines indicate the characteristic core-diameter wavenumber $k_\sigma=2\pi$ (with $\sigma=1$) and the hexagonal close packing wavenumber $k_{\rm hex}=4\pi/\sqrt{3}$, respectively.
The phase diagrams d)-f) show where each of the different maxima $k_i$ are positive $\omega(k_i)=\omega_i>0$ (the numbers $i$ in each region are the positive peak indices), while diagrams g)-i) show regions where each of the different maxima are the supremum, $\sup\{\omega_i\},\omega_i>0$. In the yellow regions, $\omega(k) \leq 0$ for all $k$.}
  \label{F1}
\end{figure}

\begin{figure}[t]
\label{PHASE_COMP_GEMC}
\centering
\includegraphics[width=\columnwidth]{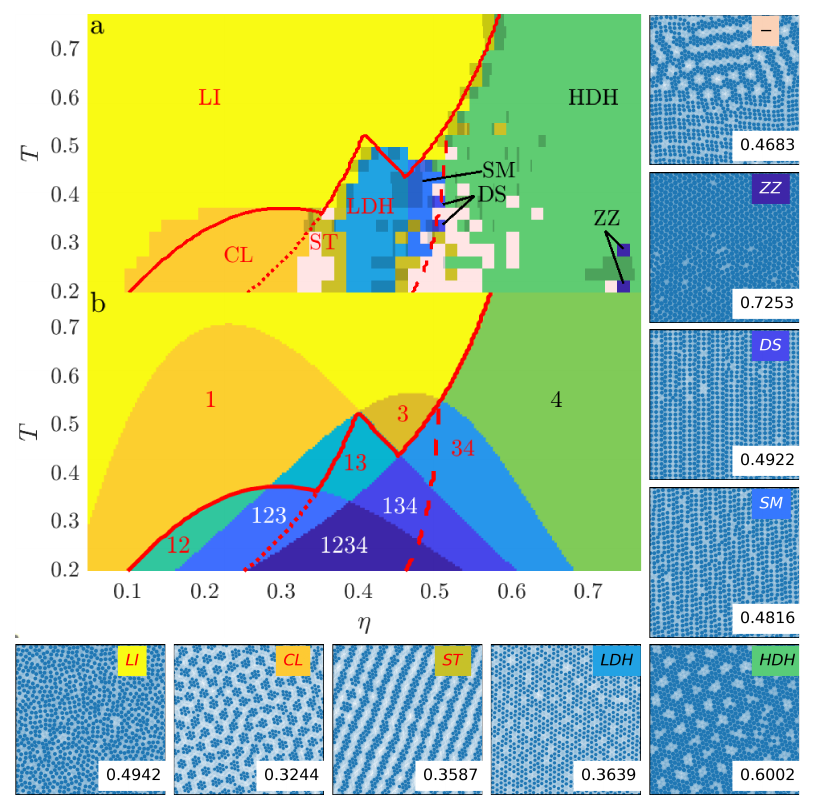}
\caption{a) Phase diagram of the HCSS system for $\lambda=3.7$ obtained via GEMC.  Each pixel represents an individual GEMC simulation. For each we plot either i) the color of the phase if both simulation boxes contain the same phase, or ii) the color of the denser phase and the extent of the coexistence region (as gray shaded area) if phase coexistence is observed or iii) plot the pixel pale pink in cases where simulations were indecisive, i.e.\ we found more than one phase in one of the simulation boxes, such as the top right example labeled `-'. The color coding corresponds to the color of the labels in the single GEMC simulation box snapshots displayed along the bottom and on the right. The naming acronyms are defined in the text, while the numbers on each are the $\eta$ values. b) The corresponding region of Fig~\ref{F1}e), showing how boundaries between the different regions (red lines) match with the phase boundaries in panel a) diagram. The red lines are either boundary lines where a second mode $\omega_i$ becomes unstable, or lines where two maxima are equal, $\omega_i=\omega_j$, $i\neq j$; i.e.\ boundaries from Fig.~\ref{F1}h). Specifically, the dotted and dashed lines are where $\omega_2=\omega_3$ and $\omega_3=\omega_4$, respectively. 
}
\label{F2}
\end{figure}

By elementary analysis of $\omega(k)$ we can identify regions in the phase diagram where the liquid is stable by determining if $\omega(k)\leq0$ for all $k$.
Consequentially, in regions where $\omega(k)>0$ for some $k$, the formation of structured phases is to be expected.
At high  $T$, on increasing $\eta$, this is precisely what we observe.
However, at lower $\eta$ (and low $T$) there is a subtlety to making this analysis with a dispersion relation $\omega(k)$ obtained from a mean-field approximate DFT:
What we actually find here is that when $\omega(k)$ exhibits only one small positive peak (i.e.\ only one weakly unstable mode) then MC simulations show that the system can in fact still be in the disordered liquid state, albeit this liquid is highly correlated, with clear signature of the favored length scale.
In other words, although the linearized mean-field functional predicts the liquid to be unstable and the solid to form, in reality thermal fluctuations are strong enough to prevent freezing in this regime.
At lower $\eta$, it is only when $\omega(k)$ exhibits at least two positive peaks that periodic (or QC) phases are found as the thermodynamic equilibrium state.
Moreover, the locus in the phase diagram where the additional $k_i$ become unstable is close to the coexistence line between the liquid and crystalline phases, as identified from the MC simulations.
Thus, the line in the phase diagram where there are multiple unstable modes $k_i$ provides a good estimate for where freezing occurs at low $\eta$.
We illustrate this in Figs.~\ref{F1} and \ref{F2}.
We believe more than one unstable mode in $\omega(k)$ is required for crystals (or QC) to be stable at low $\eta$ because at these densities the structures are rather open (with plenty of free space) and so require the coupling of periodic density modes with more than one wavelength to stabilize them.

Figure~\ref{F1}a)-c) show typical examples of $\omega(k)$ for three different shoulder ranges, $\lambda=2.5$, 3.7 and 4.9.
The shoulder height $\epsilon=1/T$ and the packing fraction $\eta$ can be identified from the position of the correspondingly colored dots in the diagrams below each in panels d)-i).
The relevant range is $k \in (0,k_{\rm hex})$, where $k_{\rm hex}=4\pi/\sqrt{3}$ is the wavenumber corresponding to hexagonal close-packing of hard discs with diameter $\sigma=1$ (vertical black dashed line). 
The other relevant scale is $k = 2 \pi/\sigma$, marked as a vertical red dashed line. 
The locations of the peaks in $\omega(k)$, denoted by the peak-value wavenumbers $k_i$ (where $i = 1, 2, 3, \dots)$, with $\omega_i = \omega(k_i)$.
These arise from an interplay between oscillations in $\omega(k)$ arising from the hard-core interactions with those due to the shoulder-interactions, having amplitude proportional to $\epsilon$.
As shown in Fig.~\ref{F1}a)--c), for $\lambda=2.5$, $\lambda=3.7$ and $\lambda=4.9$, there are three, four and five possible unstable modes $k_i$ respectively.
Note for $\lambda = 4.9$, $k_6>k_{\rm hex}$.

Following three different pathways through the $(T, \eta)$-plane, we observe the following trends: 
(i) On decreasing $T$ -- see Fig.~\ref{F1}a), d) and g) -- we find the maxima of $\omega(k)$ increasing, till they eventually become positive in value. 
(ii) Keeping $T$ fixed and increasing $\eta$ -- see Fig.~\ref{F1}b), e) and h) -- we observe a shift upwards in $\omega(k)$ at higher $k$.
For $\lambda=3.7$ this leads to the emergence of a single dominant peak at $k_4\approx k_\sigma=2\pi$, the characteristic wavenumber of the hard core.
For even higher densities, $k_4$ tends towards $k_{\rm hex}$, converging at $\eta\approx0.904$.
The increase of $\omega_4=\omega(k_4)$ is accompanied by a suppression of the growth rates $\omega_i$ for $i<4$.
This can even lead to the disappearance of peaks at smaller $k$ -- see e.g.\ for $k_1$ at $\eta=0.7$ (black line).
Note that while the $k_4$ peak shifts with varying $\eta$, the positions $k_i$ of the other peaks do not change significantly, due to the quadratic density dependency of the RPA shoulder-interaction part of $F_{\rm ex}$.
(iii) In Fig.~\ref{F1}c), f) and i) we observe a somewhat similar behavior on a diagonal path in the phase diagram for $\lambda=4.9$, with the $k_5$ peak growth rate $\omega_5=\omega(k_5)$ increasing with increasing $\eta$.

\begin{figure}[]
  \centering
  \includegraphics[width=1.\columnwidth]{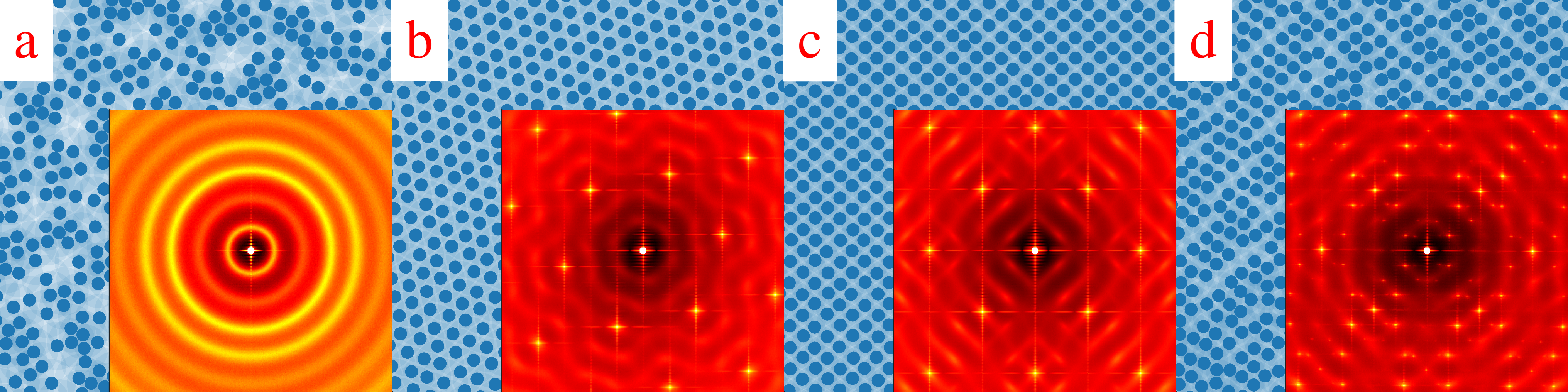}
  \includegraphics[width=1.\columnwidth]{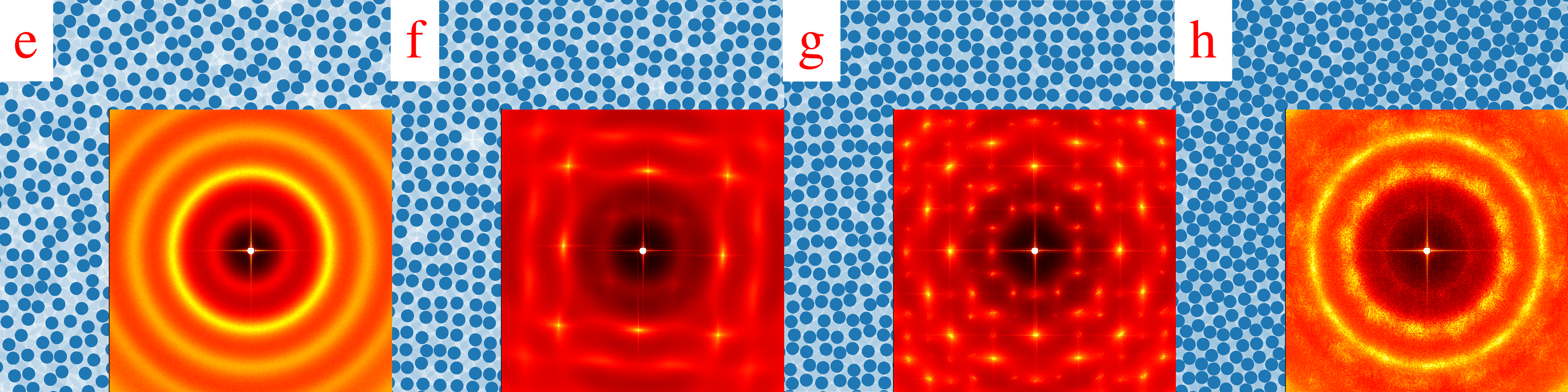}
  \caption{Snapshots of some phases encountered in GCMC simulations (in the $(x, y)$-plane), with fixed $(\mu(\eta),T,V)$. The respective insets show heatmap plots of the static structure factor $S(k_x, k_y)$, obtained via averaging along GCMC simulations, with color scale (different in each case) going from dark at low values to yellow at the highest values. Panels a)-d) are for $\lambda=3.7$; a) shows the liquid state at state point ${\cal S}=(\eta,T)=(0.35,0.4375)$, b) a low density hexagonal (LDH) crystal at ${\cal S}=(0.425,0.3751)$, c) a rhombic crystal at ${\cal S}=(0.5,0.25)$ and d) a disordered smectic phase at ${\cal S}=(0.475,0.4375)$.
  Panels e)-h) are for $\lambda=2.5$; e) shows the liquid at ${\cal S}=(0.4,0.333)$, f) a disordered low density cubic crystal at ${\cal S}=(0.45,0.333)$ with wavenumber ratio $k_3/k_2$ close to $\sqrt2$, g) a precursor to a 12-fold QC at ${\cal S}=(0.54,0.333)$ and h) a 12-fold QC at  ${\cal S}=(0.6,0.333)$.}
  \label{F3}
\end{figure}

In Fig.~\ref{F2} we revisit the $\lambda=3.7$ phase diagram (see Fig.~\ref{F1}e), complemented by snapshots of some of the phases found.
Panel a) summarizes the outcome of scanning through parameter values with Gibbs ensemble MC (GEMC) to identify the phases that arise \cite{footnote1}.
Figure~\ref{F2}b) is the corresponding region from Fig.~\ref{F1}e), showing where the $\omega_i$ of the various peaks $k_i$ are positive.
The red lines show where some of the $\omega_i$ either change sign or where $\omega_i=\omega_j$ with $i\neq j$ (see caption).
Note that these line-up rather well with the freezing boundaries.
Along the bottom and right of Fig.~\ref{F2} are examples of some of the phases observed, labeled with combinations of letters to indicate the various phases and background color corresponding to the color used in the top phase diagram.
For example, `LI' refers to the liquid phase and is colored yellow.
At lower $\eta$, structured phases occur at low $T$ and are primarily shaped by the long-wavelength $k_1$ mode (e.g.\ the cluster CL and stripe ST phases).
The region where $\omega_3$ is largest is predominantly populated by the low-density-hole LDH phase, dominated by density modulations with wavenumber $k_3$ (also see Fig.~\ref{F3}).
Along the red dashed line in Fig.~\ref{F2} the balanced competition between the $k_3$ and $k_4$ modes leads to a particularly rich phase behavior and the occurrence of several phases involving both modes, such as the striped DS and SM structures, as well as the rhombic phase observed in grand canonical MC (GCMC) simulations [see Fig.~\ref{F3}c)], not resolved with GEMC.
At higher $\eta$, we find excellent correspondence between where $\omega_4>0$ and the region of the high-density-hole HDH phase, dominated by the $k_4$ mode.
The closeness of the LI-HDH phase transition to the $(\omega_4=0)$-line highlights the predictive capabilities of the theory.
At very high densities we also encountered a zig-zag ZZ phase -- c.f.~Refs.~\cite{imperio2006microphase, Somerville2020}.
Overall, we see that our simple theory based on $\omega(k)$ can quickly predict (i) where to look for crystalline phases and (ii) what combinations of lengthscales to expect in these.
However, with our approximate DFT, MC simulations are needed to identifying exactly what crystal phases will form.

\begin{figure}[]
  \centering
  \includegraphics[width=0.65\columnwidth]{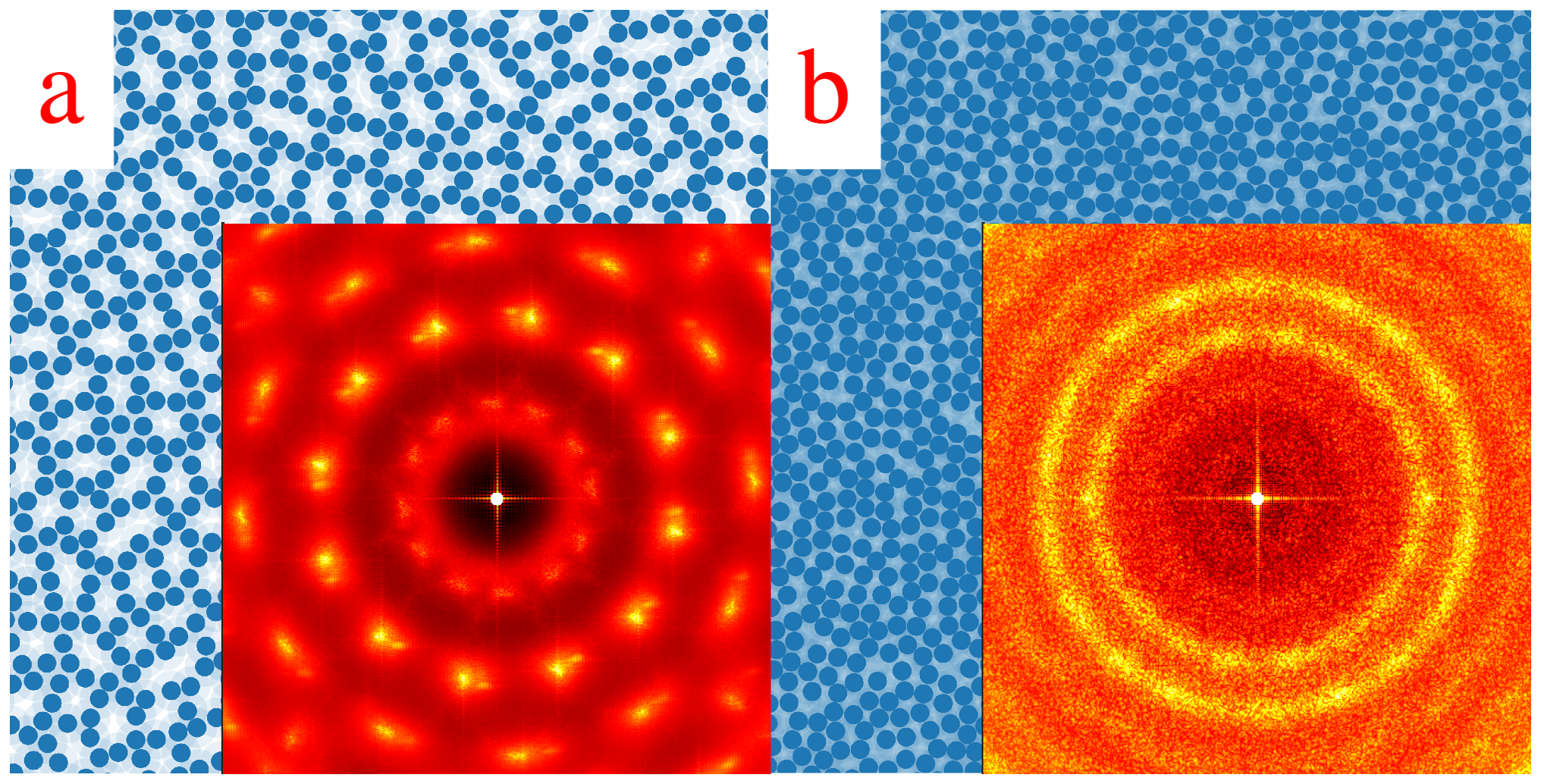}
  \caption{Left: a 12-fold QC at $(\lambda,\epsilon,\eta)=(2.1,5.81,0.35)$ with wavenumber ratios $k_2/k_1=1.9225\approx\sqrt{2+\sqrt{3}}\approx1.932$ and $k_3/k_2=1.419\approx\sqrt{2}\approx1.414$. Right: an 18-fold QC at $(\lambda,\epsilon,\eta)=(3.42,5.03,0.65)$ with $k_4/k_3=1.289\approx2\sin(2\pi/9)$.}
  \label{F4}
\end{figure}

Figure~\ref{F3} shows snapshots from GCMC simulations (single box with fixed $T$, $\mu$ and volume $V$ \cite{Allen2017, frenkel_smit}), together with the corresponding static structure factor $S({\bf k})$.
The top panels are for $\lambda=3.7$, while the bottom panels for $\lambda=2.5$, with average density increasing from left to right.
For both $\lambda$, we observe the typical ring-like (no angular dependence) static structure factor $S({\bf k}) = S(k)$ in the liquid states a) and e) with maxima at radii $k_i$.
When $\eta$ is increased and/or $T$ lowered (such as in panels a) to c) or e) to f) in Fig.~\ref{F3}), structured phases emerge, formed from the dominant density modes already present in the liquid.
They are either formed of modes with a single $k_i$, like the LDH phase in Fig.~\ref{F3}b), which essentially has the single lengthscale $2\pi/k_3$; alternatively they originate from a coupling of two modes such as in the cubic coupling of $k_2$ and $k_3$ in f) or rhombic $k_3$ and $k_4$ in c), respectively.
More complex structures including QCs come from having two unstable wavenumbers $k_i$ and $k_j$, with the requirement that corresponding wavevectors $\mathbf{k}_i$ and $\mathbf{k}_j$ add up to another already favored wavevector \cite{barkan2011stability, archer2013quasicrystalline, barkan2014controlled, archer2015soft,  savitz2018multiple, Ratliff2019, subramanian2021density, Archer2022, lifshitz1997theoretical, alexander1978should, bak1985phenomenological, rucklidge2012three, castelino2020spatiotemporal}.
This can be seen in the QC-precursor in g) and the 12-fold QC in h).
QCs have been seen previously in HCSS systems:
Refs.~\cite{Dotera2014, ziherl2016geometric, Dotera2017} present a recipe based on real-space geometric arguments for how to choose $\lambda$ and $\eta$, to ``design'' QCs, that works well at low $T$.
Here, we extend to the present HCSS system the Fourier-space approach based on $\omega(k)$ of \cite{barkan2011stability, archer2013quasicrystalline, barkan2014controlled, archer2015soft,  savitz2018multiple, Ratliff2019, subramanian2021density, Archer2022}, that was previously developed for soft potentials.
Our approach also works at higher $T$, near to melting, so nicely complements the low-$T$ method of \cite{Dotera2014, ziherl2016geometric, Dotera2017}.
To tailor QCs (as foreshadowed above) one must choose $\eta$, $T$ and $\lambda$ so that $\omega(k)$ has two unstable $k_i$.
Both must have similar growth rates $\omega_i$ and must have $k_i/k_j$ ``close'' to certain (irrational number) ratios \cite{barkan2011stability, archer2013quasicrystalline, barkan2014controlled, archer2015soft,  savitz2018multiple, Ratliff2019, subramanian2021density, Archer2022, lifshitz1997theoretical, alexander1978should, bak1985phenomenological, rucklidge2012three, castelino2020spatiotemporal} (explained in detail in the SI below).
Figure~\ref{F4} shows two outcomes of such a search, based on analysis of $\omega(k)$.
GCMC simulations were then used to confirm the predicted 12-fold and 18-fold symmetric QCs at each of these state points, respectively.

To summarize, we have presented a versatile general approach for navigating complex phase diagrams in soft matter systems.
It only requires a suitable functional $\Omega[\rho]$, from classical DFT.
Via simple analysis of the dispersion relation $\omega(k)$ (obtained directly from $\Omega[\rho]$), one can easily predict where ordered (or even QC) phases are to be expected.
The approach has been applied to 2D HCSS systems exhibiting up to 10 emerging ordered phases; note that the theory is actually easier in 3D (where the Fourier transforms are simpler), at the cost of much more laborious simulations.
Likewise, via $\omega(k)$ we can {\it design} potentials giving rise to specific complex crystal/QC structures, by choosing parameters to give peaks in $\omega(k)$ in certain ratios.
Hand-in-hand with MC simulations, our general approach is computationally cheap and works at high $T$.
It identifies at which state points to deploy the computationally expensive MC simulations, speeding up greatly the mapping of phase diagrams.

Acknowledgments: The MC computer simulation results presented here were enabled via a generous share of CPU time, offered by the Vienna Scientific Cluster (VSC) under Project No.~71263. The authors thank Ms.~Katrin Muck for her guidance related to the use of HPC. AJA~gratefully acknowledges support from the EPSRC under Grant No.~EP/P015689/1. This research was funded in part by the Austrian Science Fund (FWF) under project no. PIN8759524, gratefully acknowledged by GK. 


%

 \onecolumngrid
 \vspace{115mm}

 \pagebreak
 \begin{center}
 {\Large \bf Supplementary information for:\\
Navigating complex phase diagrams in soft matter systems}\\
\vspace{1mm}
{\bf Michael Wassermair, Gerhard Kahl, Roland Roth and Andrew J.~Archer}
 \end{center}

\setcounter{equation}{0}
\renewcommand{\theequation}{S\arabic{equation}}


\section{Free energy functional}

Here, we describe the Helmholtz free energy functional $F[\rho]$ used in our work.
The ideal gas contribution, the first term in Eq.~\eqref{eq:grand_pot} in the main text, is \cite{Evans1979, hansen13}
\begin{equation}\label{DFT3}
F_{\rm id}[\rho] = k_{\rm B} T \int \rho({\bf r}) \left[ \log
	\Lambda^2 \rho ({\bf r}) - 1 \right] d {\bf r},
\end{equation}
where $\Lambda$ is the thermal de Broglie wavelength. 
The excess Helmholtz free energy functional $F_{\rm ex}[\rho]$, is decomposed into the sum of two terms:
\begin{equation}\label{eq:split}
    F_{\rm ex}[\rho]=F_{\rm core}[\rho]+F_{\rm shoulder}[\rho].
\end{equation}
The first term incorporates the contribution due to the hard-core particle interactions and is approximated as \cite{wassermair2024fingerprints, Roth2010, Roth2012}
\begin{multline}\label{FMT_ex}
F_{\rm core} [\rho] = k_{\rm B }T \int \Bigg[ -n_0 ({\bf r}) \log{(1-n_2 ({\bf r}))}\\
+\frac{1}{4\pi(1-n_2 ({\bf r}))} \left( \frac{19}{12}({\bf n}^{(0)}({\bf r}))^2-\frac{5}{12}{\bf n}^{(1)}({\bf r})\cdot {\bf n}^{(1)}({\bf r})-\frac{7}{6}{\bf n}^{(2)}({\bf r}) \cdot {\bf n}^{(2)}({\bf r})\right) \Bigg] d{\bf r}.
\end{multline}
This is the result from FMT, built from a set of weighted densities that are either scalar 
\begin{equation}
n_\alpha ({\bf r}) = 
\left[ \rho\otimes\omega_\alpha \right] ({\bf r}), \hspace{1cm} \alpha= 0, 2,
\end{equation}
or tensorial in nature:
\begin{equation}
{\bf n}^{(m)}({\bf r}) = 
\left[ \rho\otimes{\bf \omega}^{(m)} \right] ({\bf r}), \hspace{1cm} m=0, 1, 2 .
\end{equation}
The weighted densities are defined as convolutions involving the set of weight functions \cite{Roth2012}
\begin{equation}
\omega_0(r)=\frac{\delta(R-r)}{2\pi R} ~~~~ {\rm and} ~~~~~
\omega_2(r)=\Theta(R-r),
\end{equation}
where {$R=\sigma/2=1/2$}, $\delta(r)$ is the Dirac delta and $\Theta(r)$ is the Heaviside step-function.
The tensorial weight functions are given by 
\begin{equation}
{\bf \omega}^{(m)}({\bf r})=\delta(R-|{\bf r}|)\underbrace{{\bf \hat{r}}\dots{\bf \hat{r}}}_{\text{$m$-times}}.
\end{equation}
The rank $m$ tensorial weight function arises from taking $m$ tensor products of the unit vector ${\bf \hat{r}}$ with itself.
The second term in Eq.~\eqref{eq:split} is the following RPA approximation \cite{Evans1979, hansen13, wassermair2024fingerprints}
\begin{equation}\label{DFT_RPA_fex}
F_{\rm shoulder}[\rho]=\frac{1}{2}\int\int\rho({\bf r})\rho({\bf r'})\phi_{\rm shoulder}(|{\bf r}-{\bf r'}|) d{\bf r} d{\bf r'},
\end{equation}
where
\begin{equation}\label{eq:pot_shoulder}
	\phi_{\rm shoulder}(r)=\begin{cases}
		\epsilon &0 < r <\lambda\sigma ~, \\
		0 &\lambda\sigma \leq r ~, \\
	\end{cases}
\end{equation} 
is the repulsive shoulder part of the pair potential.
Note that this has been extended inside the core of the particles.

\section{Dispersion relation $\omega(k)$}

We now move on to give details of the dispersion relation $\omega(k)$ generated by the $F[\rho]$ described above. 
In the expression for $\omega(k)$ given in Eq.~(6) in the main text, we use the following result for the Fourier transform of the pair direct correlation function, 
\begin{equation}\label{eq:c_split}
\hat{c}(k) = \hat{c}_{\rm core}(k) + \hat{c}_{\rm shoulder}(k),
\end{equation}
which is generated by the DFT given above in Eqs.~\eqref{DFT3}, \eqref{eq:split}, \eqref{FMT_ex} and \eqref{DFT_RPA_fex}.
The first term, arising from the hard-disk core repulsion treated using Eq.~\eqref{FMT_ex}, is the following \cite{wassermair2024fingerprints, Thorneywork2018}:
\begin{multline} \label{c_FMT_hc}
\hat{c}_{\rm core}(k)=\frac{\pi}{6(1-\eta)^3k^2}\bigg[-\frac{5}{4}(1-\eta)^2k^2J_0(k/2)^2+\left( 4((\eta-20)\eta+7)+\frac{5}{4}(1-\eta)^2k^2\right)J_1(k/2)^2\\+2(\eta-13)(1-\eta)k J_1(k/2)J_0(k/2)\bigg],
\end{multline}
where $J_n(x)$ are Bessel functions of order $n$.
The second term in Eq.~\eqref{eq:c_split} is the contribution from the shoulder, generated by the functional in Eq.~\eqref{DFT_RPA_fex}, and may be written as \cite{wassermair2024fingerprints}
\begin{equation} \label{phi:ss_k}
\hat{c}_{\rm shoulder}(k) = -\beta8\pi\epsilon\lambda\frac{J_1(\lambda k)}{k} .
\end{equation}
Thus, Eqs.~\eqref{eq:c_split}--\eqref{phi:ss_k} together with Eq.~\eqref{eq:omega} in the main text, 
specify the approximation for the dispersion relation used throughout in the present study.

\section{Further observations about the $k_i$ instability lines}

{In the main text, we show that periodic phases are not observed at state-points in the phase diagram that are inside the $k_1$ instability line, but outside of the other instability lines -- see Fig.~\ref{F2}.
Based on this observation, we argue in the main text that for periodic structures to occur at {\em lower densities} (where crystal structures are rather open, having plenty of free volume), the mean-field DFT must predict a second unstable mode.
It is the coupling of these two modes that stabilizes the low density crystals.
An alternative explanation for this observation is that the mean-field DFT predicts the $k_1$ instability line to be too high in the phase diagram.
However, this argument is undermined by the observation that the DFT clearly does seem to be locating the other ($k_2$, $k_3$, $k_4$, ...) instabilities in the right place in the phase diagram, as can be inferred from the fact that the red lines in Fig.~\ref{F2}a match the boundaries between different phases.
Since Fig.~\ref{F2} shows that all except the $k_1$ instabilities are essentially in the correct place, this is an argument that perhaps just the $k_1$ instability needs to be accounted for better by the DFT.}

{One possible way to improve the accuracy of the DFT that we tested, is to replace $\rho(\mathbf{r})\rho(\mathbf{r}')\to g(|\mathbf{r}-\mathbf{r}'|)\rho(\mathbf{r})\rho(\mathbf{r}')$ in the mean-field contribution to the free energy, $F_\mathrm{shoulder}$ in Eq.~\eqref{DFT_RPA_fex}, where $g(r)$ is an approximation for the radial distribution function.
See e.g.\ Refs.~\cite{Evans1979, hansen13} for a discussion of the origin of this approximation for $F_\mathrm{shoulder}$.
We assume for $g(r)$ the following very simple approximation, 
\begin{equation}\label{eq:g}
	g(r)=\begin{cases}
		0  & r\leq \sigma \\
		1+\Delta &\sigma < r <\ell, \\
		1 & r\geq\ell,\nonumber
	\end{cases}
\end{equation} 
which includes core exclusion effects for $r\leq\sigma$ and through the parameters $\Delta\geq0$ and $\ell>\sigma$ is able to include the effects of the first peak in $g(r)$. We have computer simulation results for $g(r)$ (not displayed), and from these we see that a modest, rough, but sensible choice at the state points of interest are e.g.\ $\Delta=0.5$ and $\ell=1.5$, though we also tried other values.
When using this simple approximation for $g(r)$, we find that {\em all} of the instability lines move around the phase diagram, not just the $k_1$ instability line, depending on the values of $\Delta$ and $\ell$.
Recall that the positions of the instability lines depend on the heights of the peaks in $\omega(k)$ and specifically on where these peaks have height $\omega(k_i)=0$. Since all of the peaks in $\omega(k)$ move with varying $g(r)$, not just the $k_1$ peak, this approach is clearly not the correct solution.
Instead, what is needed is something that suppresses the $k_1$ peak in $\omega(k)$, while leaving the remaining peaks essentially unchanged. In other words, something is needed that incorporates the effect of long-wavelength (smaller $k$) fluctuations in the system, but leaves the description of the smaller wavelength (larger $k$) untouched. We currently do not know a way of achieving this, but these ideas do give hints towards possible avenues for future research.}

\section{Monte-Carlo simulation methods}
\label{sec:simulations}

{In this section we give further details about our implementation of the Monte Carlo (MC) simulations.
Our results were obtained using the MC methods described in Ref.~\cite{wassermair2024fingerprints}, where further details and discussion of validation tests can be found.
For the sake of completeness we report here the essential details of our MC-simulations and describe the criteria used for phase identification used to construct the phase diagram in Fig.~\ref{F2}.}

\subsection{MC simulations in the grand-canonical ensemble}
\label{subsec:GCMC}

\begin{figure}[]
  \centering
  \includegraphics[width=1\textwidth]{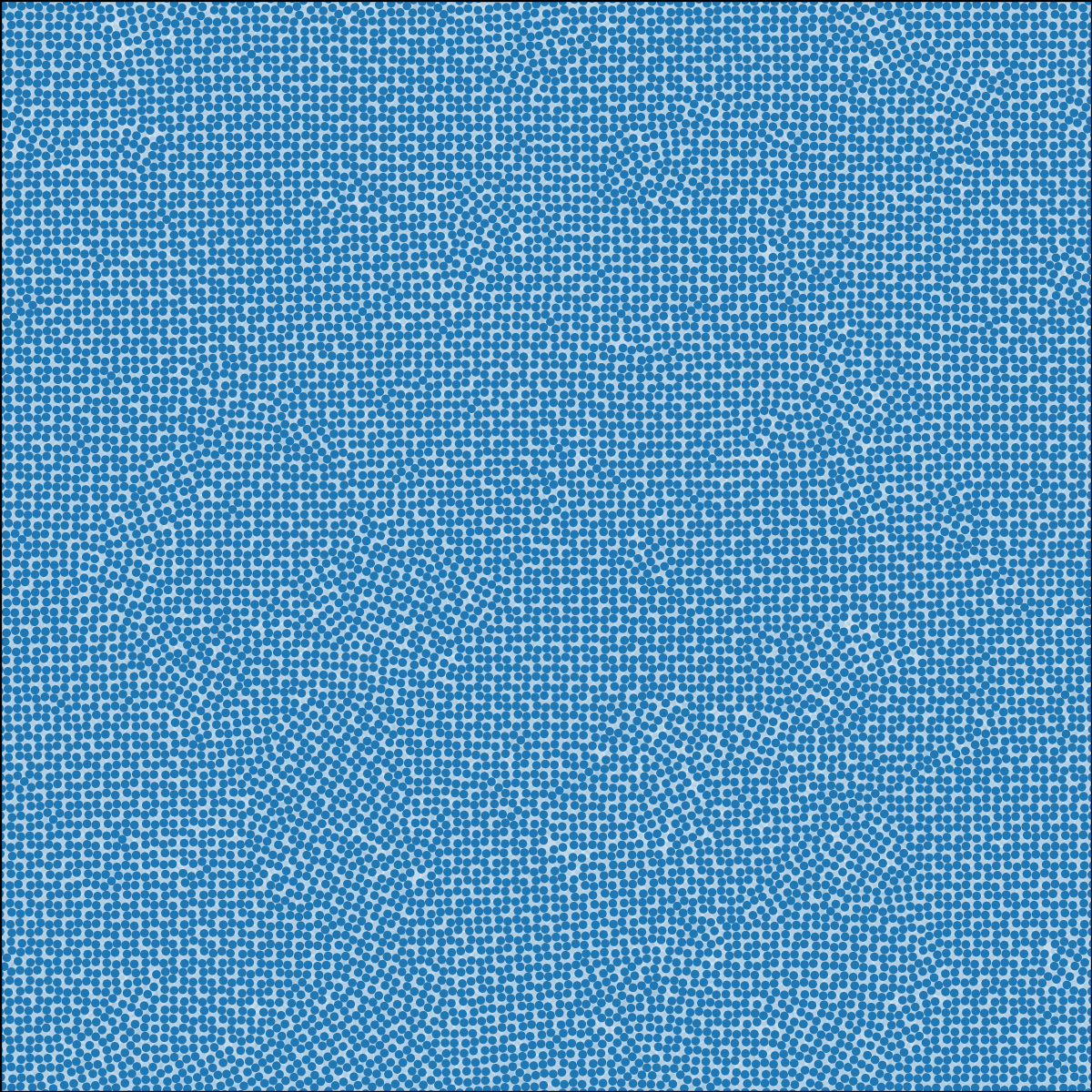}
  \caption{A snapshot showing the full extent of one of our GCMC simulations, showing a QC state obtained for $\lambda= 2.34$, $\eta = 0.653$ and $T = 0.2252$. The box size is $125\sigma\times125\sigma$ and contains around 12000 particles.}
  \label{QC23}
\end{figure}

{We have performed large scale MC simulations in the grand-canonical ensemble (GCMC), imposing particular values for the temperature $T$, the volume $V$ and the chemical potential $\mu$ of the system \cite{Allen2017, frenkel_smit}.
The number of particles $N$ can vary within the system but for practical reasons we set the maximum to be 20~000 particles in our simulations.
The ensemble is confined within a square box with periodic boundary conditions for all of our simulations.
Three types of MC ``moves'' are implemented in the course of a GCMC simulation: translational moves, particle insertions, and particle deletions, with respective probabilities $\alpha_{\rm t}$, $\alpha_{\rm i}$, and $\alpha_{\rm d}$ and suitably adapted acceptance-/rejection-criteria \cite{Allen2017,frenkel_smit}.
In our simulations we have assumed that $\alpha_{\rm t} = \alpha_{\rm i} = \alpha_{\rm d} = 1/3$ and have applied the three types of moves in a random consecutive manner.
Typical simulations extend in total over $15-20 \times 10^9$ of such MC-moves. 
Equilibration of the energy was typically observed after $4-5 \times 10^9$ MC-moves and ensemble averages, e.g. for calculation of structure factors, were performed using 250 states drawn from the end of the simulation separated by $4 \times 10^7$ MC-moves for GCMC and by $2.5 \times 10^7$ MC-moves for GEMC.}

In Fig.~\ref{QC23} we display a snapshot from one of our GCMC simulations, to show the full extent of the size of these simulations.
The snapshots in the main manuscript are small subsections taken from structures like this one.

\subsection{Gibbs ensemble MC simulations}
\label{subsec:GEMC}

{The Gibbs ensemble represents a very particular ensemble within the framework of statistical mechanics \cite{Allen2017,frenkel_smit, Panagiotopoulos1987}.
The entire system is considered in the canonical (i.e., $NVT$) ensemble, which is subdivided into two ($NPT$-) subsystems (confined in sub-boxes) which are assumed to be in phase coexistence, meaning that the two subsystems are at equal temperature $T$, equal pressure $P$, and equal chemical potential $\mu$.
This is achieved by introducing three types of MC ``moves'': translational moves of particles inside each of the boxes, particle exchange between the two boxes and volume exchange between the two subsystems, while maintaining the total volume of the system.
The acceptance/rejection criteria for each type of move can be found in the Refs.~\cite{Allen2017,frenkel_smit}.
Periodic boundary conditions are applied to each sub-box.}

{In our simulation we start from two square sub-boxes (indices `1' and `2') with $N_1 = N_2 =1000$ particles, and the total number of particles, $N_1+N_2 = 2000$ being fixed. The initial box size is chosen such that $V_1=V_2=V/2$, where the total volume $V=N/\rho_{\rm b}$, where $\rho_{\rm b}$ is the starting total average density and $N=N_1+N_2$. The positions of the particles are initialized in hexagonal lattices.} 

{The simulations are performed via consecutive blocks where each block represents a sequence of the following consecutive steps: (i) $n_{\rm trans}=200$ attempts for translational MC-moves of randomly chosen particles, i.e., 100 moves per sub-box; (ii) a single $n_{\rm volume}=1$ volume change, changing the volume of one square-shaped sub-box by $\pm\Delta V$ at the cost of the other, where $\Delta V$ is drawn uniformly from the range $\Delta V\in[0,4]$, and (iii) $n_{\rm swap}=500$ attempts to swap particles from one sub-box to the other, where in each step the box in which a particle is deleted is selected randomly.
The position of a newly created particle is chosen uniformly from within the hosting sub-box; the move is immediately rejected if an overlap of hard cores with the existing particles occurs. The simulation is essentially a sequence of repeated blocks, being continued until a total number of MC-move attempts $n_{\rm moves}=n_{\rm blocks}\times(n_{\rm trans}+n_{\rm volume}+n_{\rm swap})\geq8\times10^9$ is attained. For the ensemble averages (e.g., to calculate the structure factor -- see below) each box is treated as a $\mu VT$-ensemble of its own, analogous to the GCMC calculations.} 

\subsection{Phase diagram}

\begin{figure}[]
  \centering
  \includegraphics[width=1.0\textwidth]{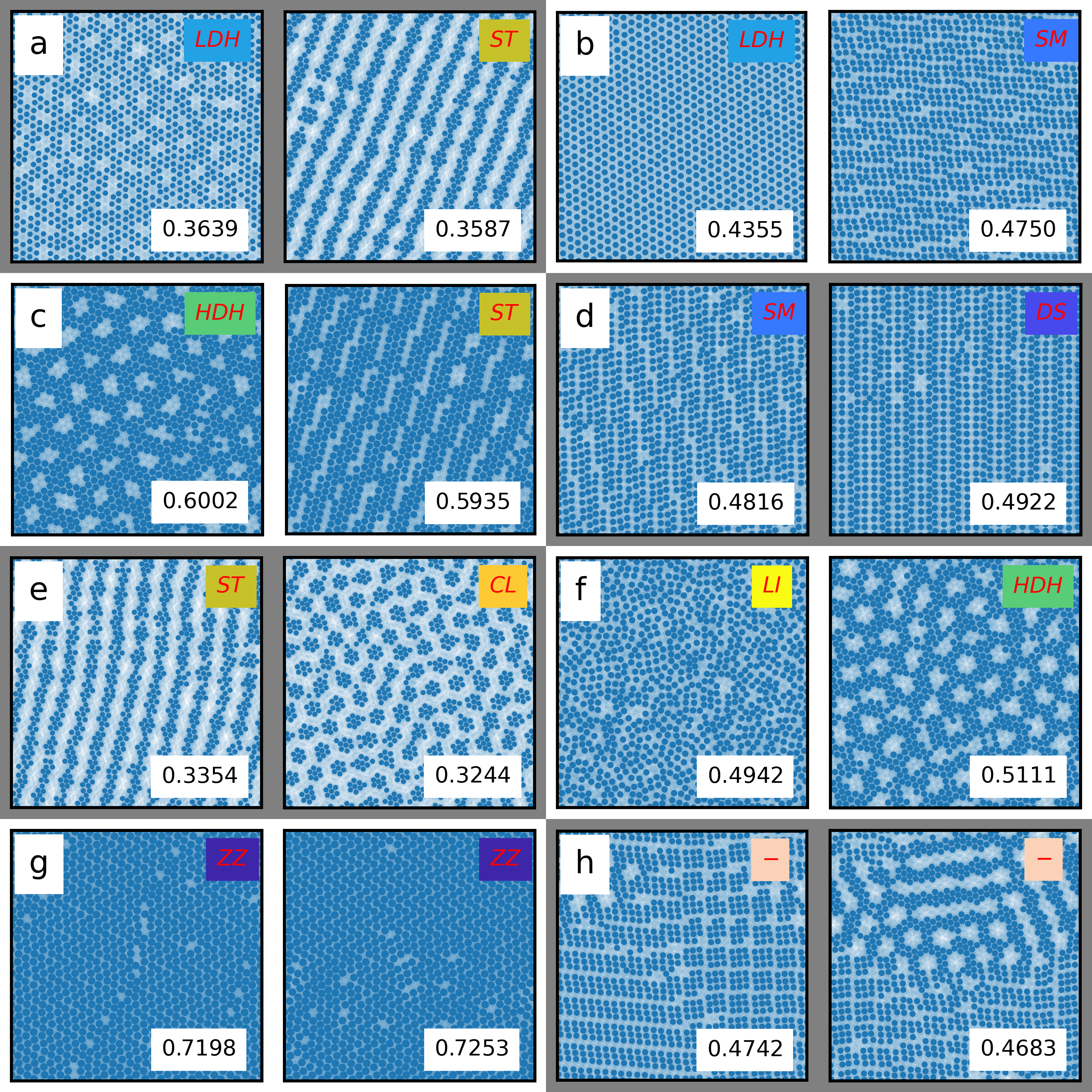}

  \caption{Further examples of phases obtained from direct simulation of phase coexistence using GEMC for $\lambda=3.7$, to supplement Fig.~\ref{F2} of the main text. The shaded or white background indicate pairs of boxes in a Gibbs-ensemble. Simulations where performed for various state points: the number on each plot is the corresponding value of $\rho_{\rm b}$. Colored letters indicate the assignment to different regions in the phase diagram in Fig.~\ref{F2} of the main text.}
  \label{FLS37}
\end{figure}

\begin{figure}[]
  \centering
  \includegraphics[width=0.5\textwidth]{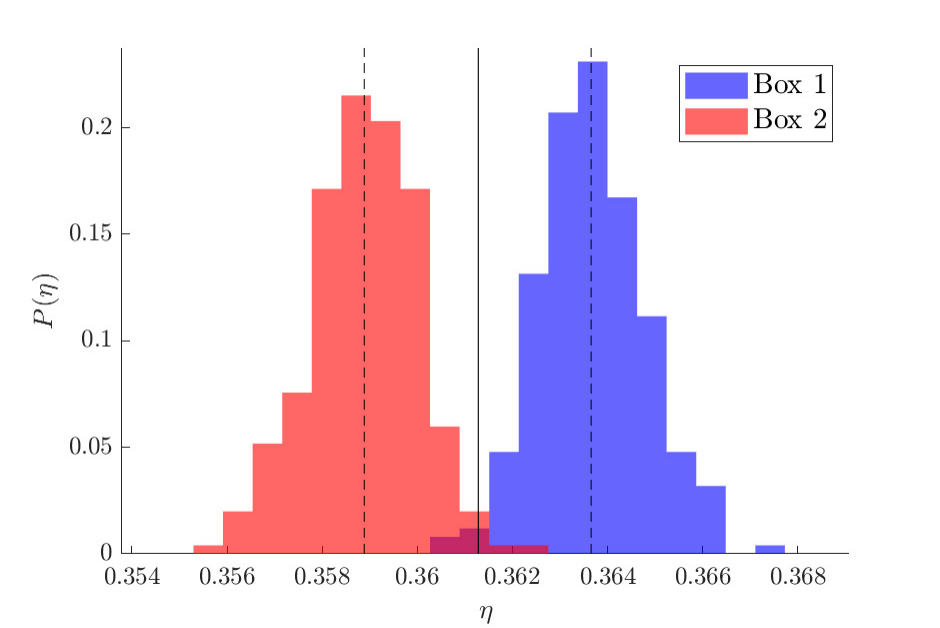}\includegraphics[width=0.5\textwidth]{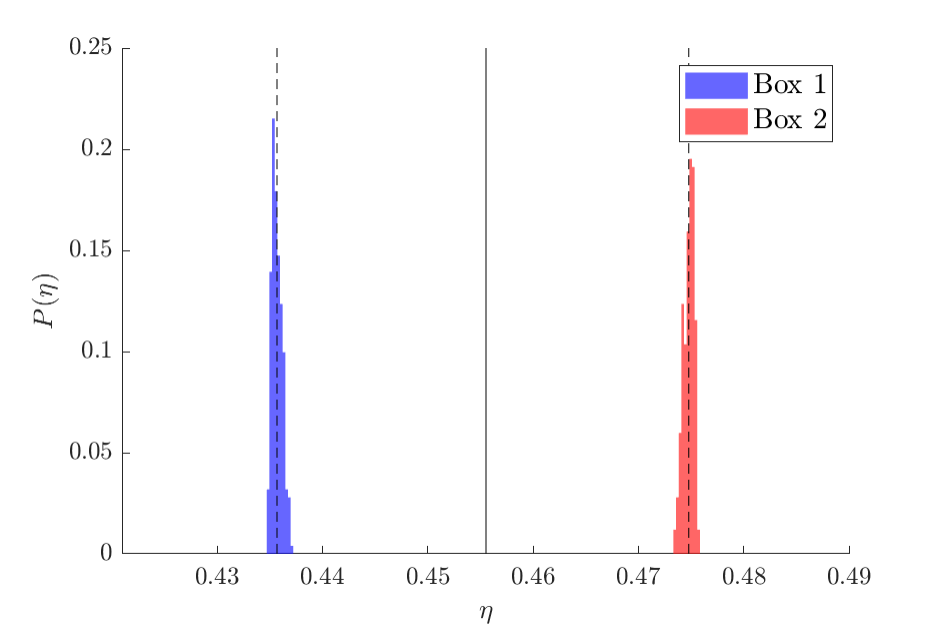}
  \caption{{Density statistics for the two GEMC simulations at $\lambda=3.7$ presented in Fig.~\ref{FLS37}. The left plot is for $\eta=0.3613,\beta\epsilon=3.077$ and the right is for $\eta=0.4555,\beta\epsilon=2.667$. Both display distinct peaks, indicating clean separation of the two phases at coexistence into the two boxes. } }
  \label{GEMC_DS}
\end{figure}

{To map out the phase diagram in Fig.~\ref{F2} of the main manuscript, the identification of the different phases in the GEMC simulations was done manually in each case, by comparison of the states observed.
Examples of some of the different phases observed are in Fig.~\ref{F2}.
To supplement this, in Fig.~\ref{FLS37} we display further examples of the various phases obtained from direct simulation of phase coexistence using GEMC for $\lambda=3.7$.
This illustrates further quite how many different phases are in the phase diagram for this value of $\lambda$.
Observing two clearly distinct pure phases in a pair of boxes such as those in Fig.~\ref{FLS37}, allows to identify clear phase coexistence.
However, any system showing more than one phase in one of the simulation boxes was marked as inconclusive (labeled `-').
Being very restrictive in this selection, leads to a high number of excluded simulations, especially at the boundaries of phases with very similar densities, as e.g.\ cluster/stripes.
If coexistence was detected (two distinct phases in the two simulation boxes), we verified convergence by ensuring two distinct Gaussian peaks in the density statistics of the system (again using averages of 250 states), as illustrated in Fig.~\ref{GEMC_DS}.
For some simulations where a visual distinction of phases was difficult, e.g.\ the liquid-cluster transition at low densities, we used the 2D structure factor to identify the phase of the states present and whether rotational symmetry was broken or not.
Although this manual phase identification was laborious, we are confident that it ultimately provides higher quality, compared to any automated phase detection scheme we could have deployed.}

\section{Finding quasicrystals}

{As mentioned in the main text, QCs come from having two unstable wavenumbers $k_i$ and $k_j$, with the requirement that corresponding wavevectors $\mathbf{k}_i$ and $\mathbf{k}_j$ add up to another already favored wavevector.
These ideas come from Refs.~\cite{bak1985phenomenological, lifshitz1997theoretical, rucklidge2012three, castelino2020spatiotemporal, barkan2011stability, archer2013quasicrystalline, barkan2014controlled, archer2015soft,  savitz2018multiple, Ratliff2019, subramanian2021density, Archer2022},
some of which are about soft matter QCs, while others are about finding quasipatterns in Faraday waves (analogue of the QC phase).
We now give a few more details about how this adding-up (or coupling) of different modes works and can be used to design QCs.}

{Mode-coupling can be understood from constructive addition of wavevectors in the 2D $k_x$-$k_y$-plane (or the corresponding volume in 3D).
The maxima of $\omega(k)$, which are located at $k_i$, $i=1,2,3,\dots$, correspond to density modes with wavevector $\mathbf{k}$ with a growth rate $\omega_i=\omega(k_i)$ that is larger than that of other nearby modes.
In other words, these are the modes that are energetically favored by the system.
In the $k_x$-$k_y$-plane, all the wavevectors with magnitude $|\mathbf{k}|=k_i$ correspond to circles (spheres in 3D) with radius $k_i$.
When we have two circles of favored modes $k_i$ and $k_j$ with $k_i<k_j$, consider e.g.\ two wavevectors $\mathbf{k}_i^1$ and $\mathbf{k}_i^2$ on the $k_i$ circle.
If this pair $\textit{add up}$ to give a vector $\mathbf{k}_j^1$ on the $k_j$ circle, this produces an energetically favorable combination, i.e.\ a favorable ``coupling'' of these density-modes.
By choosing pair potential and state point parameters so that the radii $k_i$ and $k_j$ of two of the circles have the ratio $\gamma=\frac{k_j}{k_i}$ with certain values, allows us to determine the angle between pairs of vectors on one circle that add up to a vector on the other circle.
Through our control of these angles, we can bias the system towards having different symmetries, using specific combinations wavevectors that are mutually stabilizing or ``adding up''.
This allows to make quantitative predictions of which $\textit{magnitude}$ $k_i$ of wavevectors will potentially play a role in the formation of a structured phases and further where we can expect phases that involve $\textit{combinations}$ of density-modes.
In the following section we show how we use these predictions in combination with pattern-formation considerations in order to guide the search for a quasicrystals with specific rotational symmetry.
This allows us to narrow-down significantly where to look in the vast and hugely complex phase diagram arising from the comparably simple HCSS pair interaction.}

\subsection{Finding the right ratio}
 
{In order to obtain a 2D quasicrystal of a specific discrete rotational symmetry $Q$, e.g. $Q=12$ for a 12-fold QC, we need the maxima of the dispersion relation $\omega(k)$ to be located so as to generate two circles $k_i$ and $k_j$ with exactly $Q$ discrete and evenly spaced wavevectors $\mathbf{k}_i^q$ and $\mathbf{k}_j^q$ with  $q\in(1,Q)$. Further, in order for density modes with wavenumbers $k_i$ and $k_j$ to couple, we require that two wavevectors at $k_i$ add up to a wavevector at $k_j$, i.e.\ $\mathbf{k}_i^l+\mathbf{k}_i^m=\mathbf{k}_j^n$ with $l\neq m$.
Assuming that the wavevectors at $k_j$ have an angular offset of $\frac{1}{2}\frac{2\pi}{Q}$ relative to the ones at $k_i$, we can combine these requirements into the following equation:
\begin{equation}\label{SI_QC1}
\underbrace{k_i e^{i 2\pi \frac{l}{Q}}}_{\mathbf{k}_i^l}+\underbrace{k_i e^{i 2\pi\frac{m}{Q}}}_{\mathbf{k}_i^m} = \underbrace{k_j e^{i 2 \pi \frac{n}{Q}+i\frac{1}{2}\frac{2\pi}{Q}}}_{\mathbf{k}_j^n}.
\end{equation}
Introducing the wavevector ratio $\gamma=\frac{k_j}{k_i}$ and making the arbitrary (but inconsequential) choice $n=Q$, we can rewrite Eq.~\eqref{SI_QC1} as
\begin{equation}
    \gamma=e^{i 2\pi (\frac{l}{Q}-\frac{1}{2Q})}+e^{i 2\pi (\frac{m}{Q}-\frac{1}{2Q})}.
\end{equation}
Demanding that $\gamma$ be real, we can solve the equation $\mathrm{Im}\{\gamma\}=0$ for the integers $l,m$ assuming $\gamma>1$ ($k_j>k_i$) for a given rotational symmetry $Q$.
This is easily done using the software MATHEMATICA \cite{Mathematica}.
Insertion of these solutions for $l,m$ into $\mathrm{Re}\{\gamma\}$ results in obtaining the values of $\gamma$ for which a QC with rotational symmetry $Q$ may arise from coupling of the modes $k_i$ and $k_j$.
Choosing $Q=12$ we obtain the ratios $\gamma_{\rm QC12}^1=\sqrt{2}\approx1.4142$ and $\gamma_{\rm QC12}^2=\sqrt{2+\sqrt{3}}\approx1.9319$ and for $Q=18$
the corresponding ratios are $\gamma_{\rm QC18}^1=2\sin\left(\frac{2\pi}{9}\right)\approx1.2856$, $\gamma_{\rm QC18}^2=2\cos\left(\frac{\pi}{18}\right)\approx1.9696$ and $\gamma_{\rm QC18}^3=\sqrt{3}\approx1.7321$.
Note, that these types of Fourier space considerations are not limited to QCs, but alo apply to any crystal structure involving multiple length scales.
The symmetries of the crystal will inevitably  enforce certain ratios of wavevector magnitudes.
Below we outline how we use $\omega(k)$ to find phases that are formed by wavevectors with a specific $\gamma$.}

\subsection{Estimating systematic error for $\gamma$-predictions}

\begin{figure}[]
  \centering
  \includegraphics[width=0.9\textwidth]{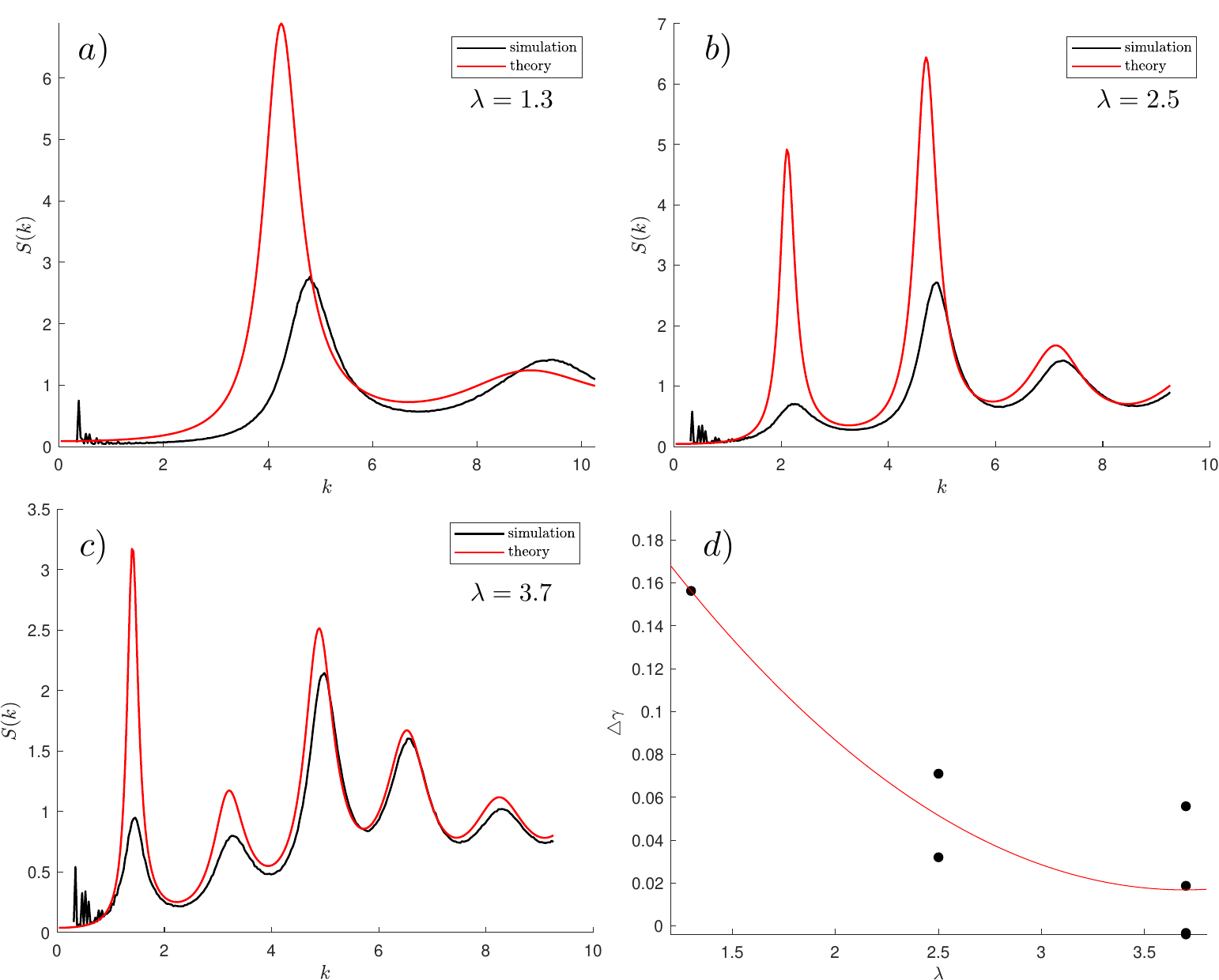}
  \caption{{Structure factors $S(k)$ obtained from GCMC simulations, compared with the corresponding analytic DFT results. These are for liquid states with $\eta=0.3$ and for $(\lambda,T)$: a) (1.3,0.273), b) (2.5,0.399), c) (3.7,0.740). In d) we display the estimated $\lambda$ dependency of the error in the $\gamma$ predicted by the DFT. The red line is a fit to a quadratic polynomial that is used later to correct the target ratio. }}
  \label{F_EST}
\end{figure}

{The aim is to find a set of parameters ${\lambda,\eta,\epsilon}$ such that the system has $\gamma=\frac{k_j}{k_i}$ for the dominant density modes $k_i$ and $k_j$ with the desired value.
In principle, the approach described in the preceding subsection is all that is needed.
However, because we only have an approximate DFT, this introduces a source of error that must be accounted for.
Luckily, the error in $\gamma$ introduced by the inaccuracy of the predicted $k_i$ and $k_j$ is rather small, very systematic and can be estimated by simply comparing the structure factor $S(k)$ of a single/few GCMC simulation in the near-by liquid state with the DFT prediction of $S(k)$.
In Fig.~\ref{F_EST} and Table~\ref{T_EST} we show examples from three simulations for $\lambda=2.5$, 3.7 and 4.9, comparing $S(k)$ between theory and simulation.
Figure~\ref{F_EST} shows that there is fairly good agreement of the positions of the peaks in $S(k)$ between the theory and simulation.
In Table~\ref{T_EST} we compare the obtained peak $k_i$ values.
Note that we denote the ratio of (consecutive) peaks as $\gamma_{ji}=\frac{k_{j}}{k_i}$.
We observe that in the limit where $\lambda$ becomes large, the error becomes rather small.
This is to be expected from our mean-field DFT treatment of the soft-shoulder interactions.
For shorter ranged $\lambda$, there is a systematic underestimation of the $k_i$, leading to an overestimation of $\gamma$.
We can estimate the $\lambda$-dependent error $\gamma_{\text{error}}(\lambda)$ in the value of $\gamma$ predicted by the DFT, by fitting a quadratic polynomial to the small number of compared peak positions in Table~\ref{T_EST}.
We subsequently use $\gamma_{\text{error}}(\lambda)$ to adjust between the desired and actual $\gamma_{ji}$ values in our search.}

\begin{table}[t]
\centering
\begin{subtable}
\centering
\begin{tabular}{c|c|c|c}
$\lambda=1.3$ & simulation & theory & $\Delta$ \\
\hline
$k_1$ & 4.801 & 4.253 &  \\
$k_2$ & 9.432 & 9.020 &  \\
$\gamma_{21}$ & 1.9646 & 2.1209 & 0.1563 \\
\hline
$\lambda=2.5$ & simulation & theory & $\Delta$ \\
\hline
$k_1$ & 2.260 & 2.105 &  \\
$k_2$ & 4.891 & 4.705 &  \\
$k_3$ & 7.243 & 7.119 &  \\
$\gamma_{21}$ & 1.4809 & 1.5131 & 0.032 \\
$\gamma_{32}$ & 2.1642 & 2.2352 & 0.071 \\
\end{tabular}
\end{subtable}
\hfill
\begin{subtable}
\centering
\begin{tabular}{c|c|c|c}
$\lambda=3.7$ & simulation & theory & $\Delta$ \\
\hline
$k_1$ & 1.455 & 1.393 &  \\
$k_2$ & 3.281 & 3.219 &  \\
$k_3$ & 4.984 & 4.891 &  \\
$k_4$ & 6.562 & 6.531 &  \\
$k_5$ & 8.295 & 8.234 &  \\

$\gamma_{21}$ & 2.2550 & 2.3109 & 0.0558 \\
$\gamma_{32}$ & 1.5234 & 1.5194 & -0.0040 \\
$\gamma_{43}$ & 1.3166 & 1.3353 & 0.0187 \\
$\gamma_{54}$ & 1.2641 & 1.2608 & -0.0033 \\
\end{tabular}
\end{subtable}
\caption{{Dominant density modes determined by the peak positions in the
$k\in(0,9)$ interval for the states in Fig.~\ref{F_EST}. 
$\Delta$ denotes the error in the prediction of the ratio $\gamma_{ji}$ 
of two consecutive peaks.}}
\label{T_EST}
\end{table}

\subsection{Pre-screening the $\lambda$-range}

\begin{figure}[]
  \centering
  \includegraphics[width=0.8\textwidth]{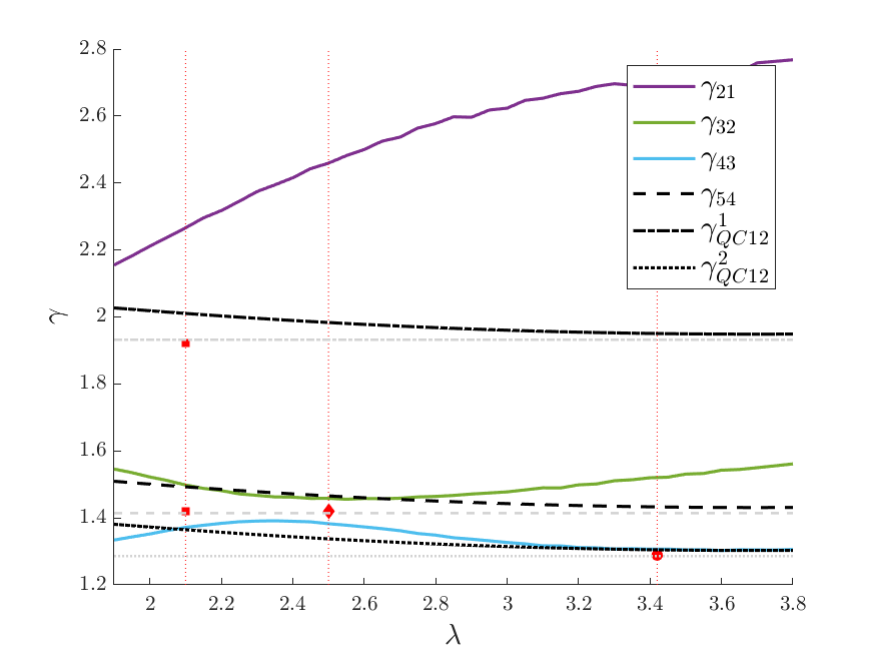}
  \caption{{Ratios $\gamma_{ji}=\frac{k_j}{k_i}$ of consecutive peaks ($k_j>k_i$) in the dispersion relation $\omega(k)$ up to the 4th peak for states at $\epsilon=\frac{13}{\lambda^2}$ and $\eta=0.5$. $\gamma$ of interest for QC formation corrected with the estimated error $\gamma_{\text{error}}$ are displayed with dashed and dash-dotted lines. The grey horizontal lines indicate the actual $\gamma_{\rm QC}$. The wavevector ratios found in the QCs obtained via GCMC simulation are depicted as red symbols and are referring to $\gamma_{21}$ and $\gamma_{32}$ at $\lambda=2.1$, $\gamma_{32}$ at $\lambda=2.5$ and $\gamma_{43}$ at $\lambda=3.42$.} }
  \label{QCgammas}
\end{figure}

{To narrow down the $\lambda$-ranges of interest for QC formation, we choose the shoulder potential height to be $\epsilon=\frac{\tilde{\epsilon}}{\lambda^2}$, with $\tilde{\epsilon}=13$, and choose $\eta=0.5$.
In other words, we make sure to be in a temperature and density regime where solid phases should arise.
At these state points, from the resulting $\omega(k)$ we compute the ratios of consecutive peaks for the first four $k_i$.
Different choices of $\tilde{\epsilon}$ only marginally altered the results displayed in Fig.~\ref{QCgammas}, where we compare the resulting $\gamma_{ji}$ values to the error corrected target values (black dashed lines) $\gamma_{\rm QC}+\gamma_{\text{error}}$ that are needed for QC formation.
From this comparison we find that density modes $k_3$ and $k_2$ have a ratio close to $\gamma_{\rm QC12}^1$ for $\lambda\approx2.1 - 2.6$, while $\gamma_{43}$ approaches $\gamma_{\rm QC18}^1$ for $\lambda>3.2$.
Note, that we also see $\gamma_{43}\approx\gamma_{\rm QC18}^1$ for $\lambda\approx2.1$ but for small $\lambda$, the involved $k_4$ is significantly bigger than $k_\mathrm{hex}=4\pi/\sqrt{3}$, the wavevector associated with a closest hexagonal packing, so this one is ignored.}

\subsection{Finding phases with coupling density modes}

\begin{figure}[]
  \centering
  \includegraphics[width=1\textwidth]{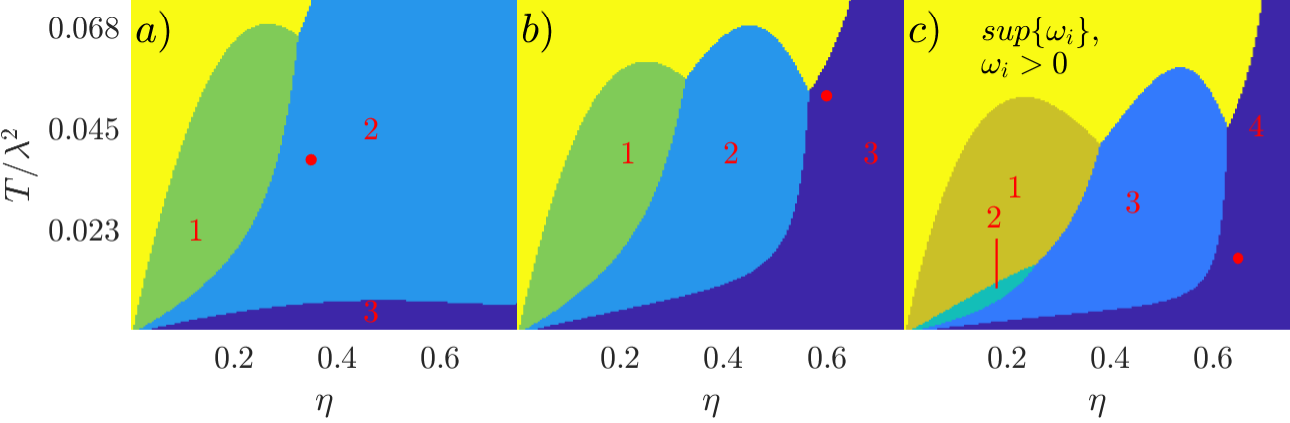}
   
  \caption{{Regions in the phase diagram where one density mode is dominant and the dispersion relation is positive. a) $\lambda=2.1$, $\lambda=2.5$ and c) $\lambda=3.42$. The red dots denote the positions where we performed the GCMC simulations in Fig.~\ref{QCex}.}}
  \label{QCPD}
\end{figure}

\begin{figure}[]
  \centering
  \includegraphics[width=0.33\textwidth]{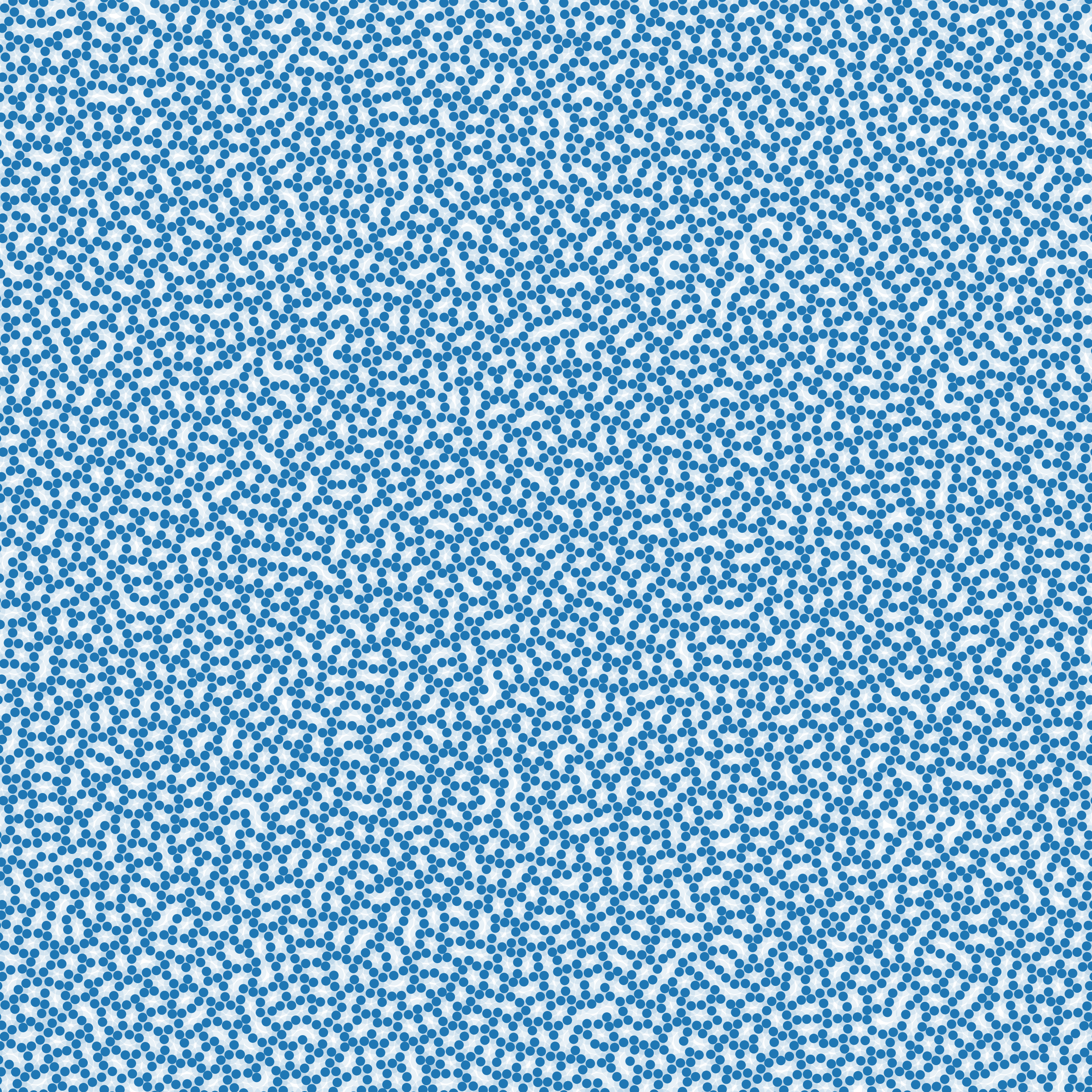}\includegraphics[width=0.33\textwidth]{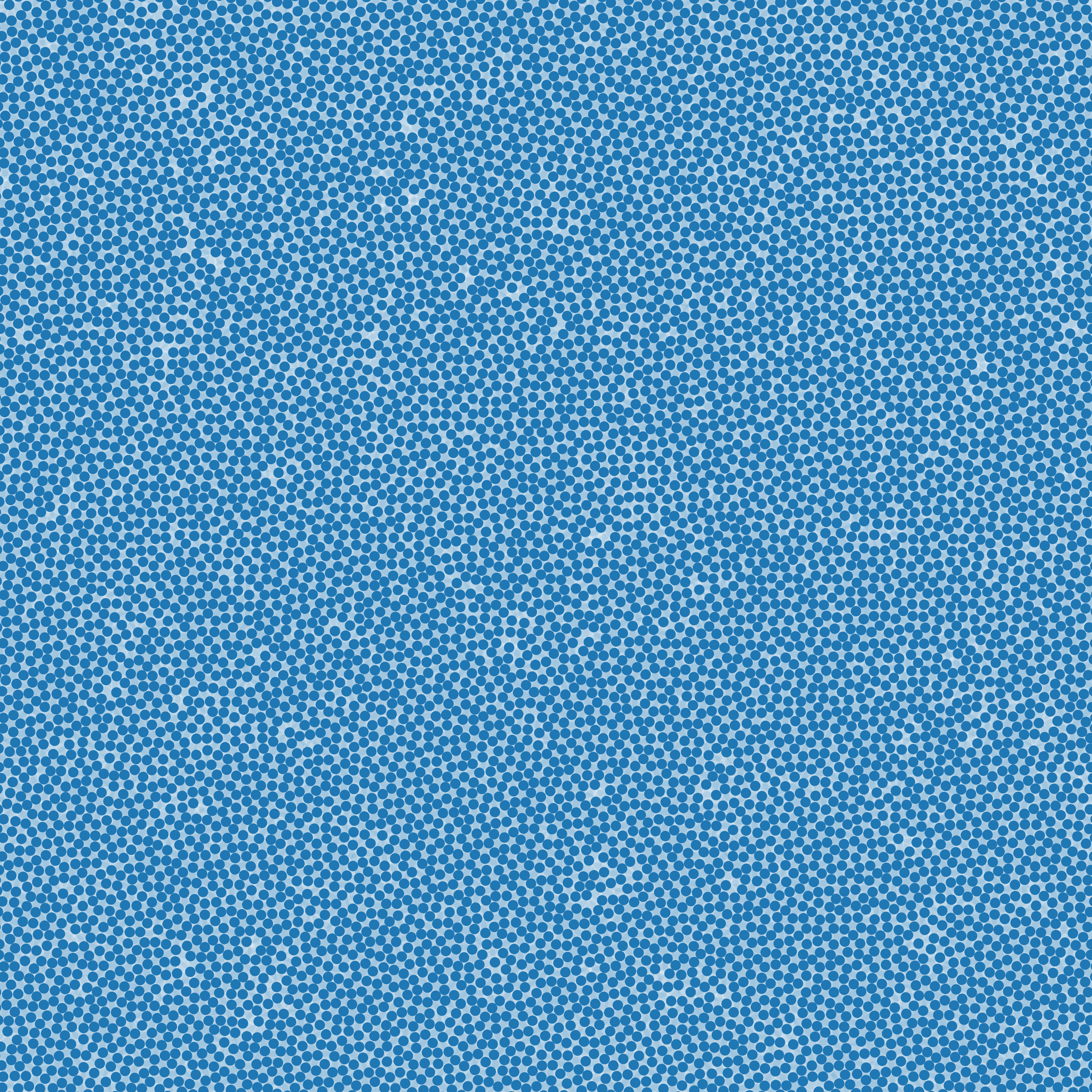}\includegraphics[width=0.33\textwidth]{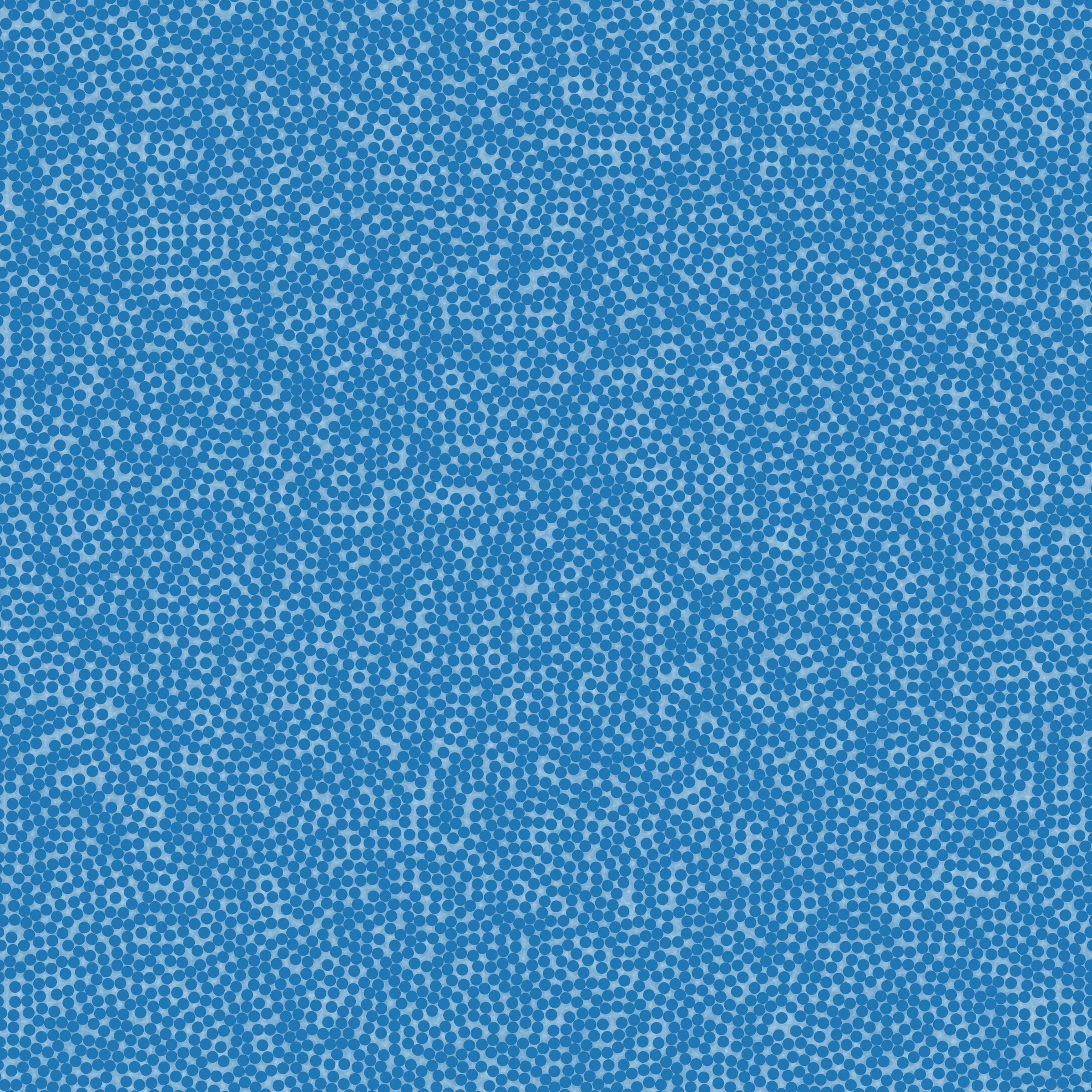}
  
  \caption{{Full extent of the GCMC simulations of QCs at (from left to right) $\lambda=2.1$, $\lambda=2.5$ and $\lambda=3.42$. Box sizes were $L=84\sigma$, $L=104.5\sigma$ and $L=92.5\sigma$, respectively.} }
  \label{QCex}
\end{figure}

{As mentioned in the main text, we expect the observed phases to be those that are formed by a combination of density modes $k_i$ and $k_j$ in regions of the phase diagram where $\omega(k_i)=\omega(k_j)$.
This condition corresponds exactly to the boundaries between colored, enumerated regions in the diagrams displayed in Fig.~\ref{QCPD} (see also Figs.~1 and 2 of the main manuscript).
Hence, after picking $\lambda$ values that have QC-compatible $\gamma_{ji}$ in the pre-screening, we performed a small number of GCMC simulations in the direct vicinity of the boundaries associated with these wavevector combinations.
In all three cases we found QCs formed by the predicted wavevector combinations and $\gamma_{\rm QC}$ close to said boundaries.
The avoidance of coexistence regions with other phases could require some small parameter changes, in order to obtain pure phase simulations (see Fig.~\ref{QCex}) as for e.g.\ $\lambda=3.42$ (original guess $\lambda=3.4$).}

\begin{figure}[]
  \centering
  \includegraphics[width=1\textwidth]{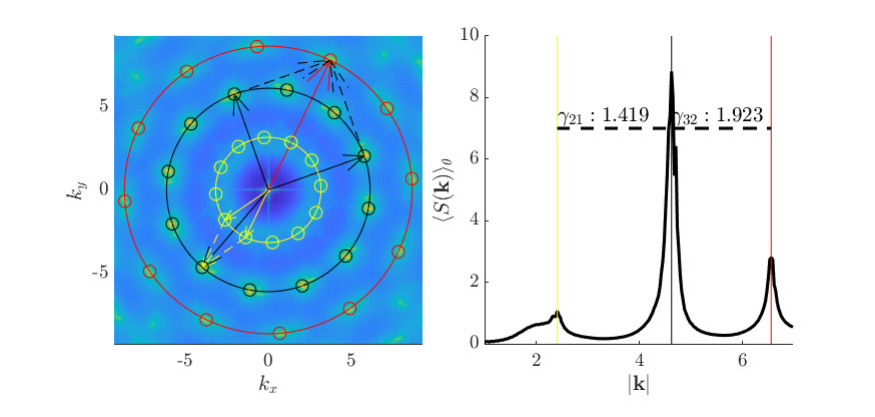}
  \caption{{Left: 2D structure factor of QC at $\lambda=2.1$ obtained from GCMC simulations. The coloured arrows (and small circles) exemplify the different wavevectors taking part in the formation of the crystal. Note, that the only matching of the overlaid diagram with the simulation structure factor was for a single peak position at $k_2$ (small black circles), whereas the positions of all yellow and red circles are derived directly from the solution of Eq.~\eqref{SI_QC1} for $\gamma_{\rm QC12}^1$ and $\gamma_{\rm QC12}^2$, showing near perfect coincidence with the simulation. Right: The polar radial average of the the structure factor showing 3 distinct peaks at $k_i$, forming ratios very close to the analytically predicted $\gamma_{\rm QC12}^1\approx1.414$ and $\gamma_{\rm QC12}^2\approx1.932$.  } }
  \label{QCsk}
\end{figure}

{A particularly interesting case is the QC at $\lambda=2.1$, where we found QCs both close to the boundary where $\omega_3=\omega_2$, but also at significantly higher temperatures -- see Fig.~\ref{QCPD} a).
We believe the reason for this very high stability is that there is in fact coupling of {\em three} density modes, as illustrated in the analysis of the structure factor in Fig.~\ref{QCsk}.
While $k_3$ and $k_2$ couple, having the ratio $\gamma_{\rm QC12}^1\approx1.414$, $k_2$ and $k_1$ also couple, having the ratio $\gamma_{\rm QC12}^2\approx1.932$.
This leads to a QC with three distinct length scales, which is stable at relatively low densities and high temperatures.
}

\begin{figure}[]
  \centering
  \includegraphics[width=1\textwidth]{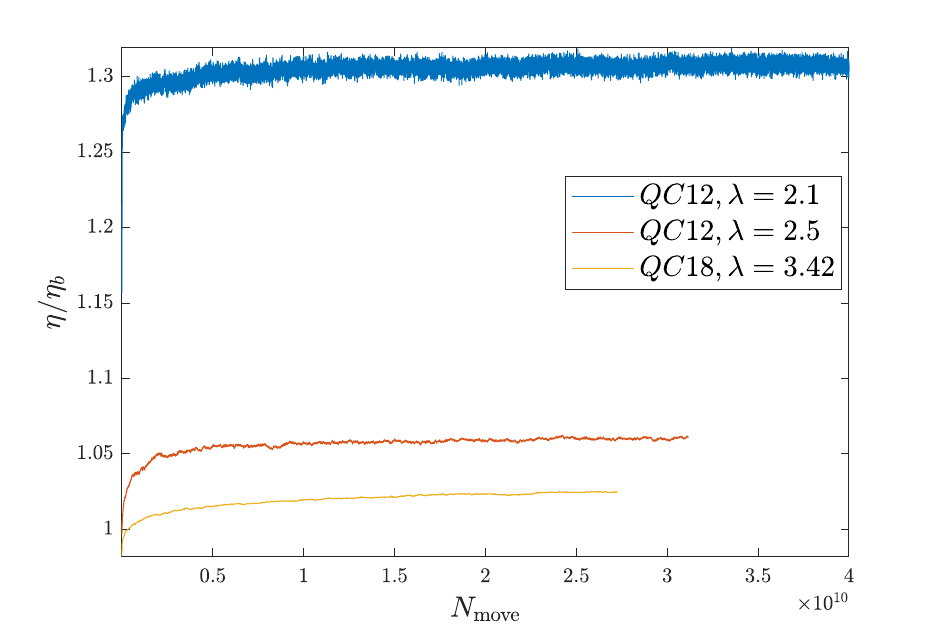}
  \caption{{Equilibration of the mean packing fraction in the course of the GCMC simulation. Normalization with the expected value of a bulk liquid with same chemical potential $\mu(\eta_{\text{b}})$ reveals a $30\%$ density increase when the liquid is crystallizing to the QC phase at $\lambda=2.1$. All systems were simulated for at least $25\times10^9$ MC-steps. }}
  \label{fig:QCDE}
\end{figure}

{Finally, we display in Fig.~\ref{fig:QCDE} an example of the equilibration towards a constant mean density in the GCMC simulations.
Comparing the equilibrated density of the QC at $\lambda=2.1$ to the density of a bulk liquid with the same chemical potential, we see a $30\%$ increase, that correlates with the observed high thermal stability.
All systems were equilibrated for more then $25\times10^9$ MC steps, leading to good equilibration of the QCs.
One could probably reduce the number of defects in the QCs (as seen in Fig.~\ref{QCex}) by optimization of the box size \cite{Archer2022} to minimize frustration, which to some extent was already done.}

\end{document}